\documentclass[english,superscriptaddress,showpacs,showkeys]{revtex4-2}
\usepackage[T1]{fontenc}
\usepackage{tikz}
\usepackage[compat=1.1.0]{tikz-feynman}
\usepackage[utf8]{inputenc}
\usepackage{color}
\usepackage{babel}
\usepackage{array}
\usepackage{units}
\usepackage{bbold}
\usepackage{amsmath}
\usepackage{amssymb}
\usepackage{graphicx}
\usepackage{bm}
\usepackage{subcaption}
\usepackage{todonotes}
\usepackage{tikz}
\usetikzlibrary{arrows.meta}

\usepackage[pdfusetitle,
bookmarks=true,bookmarksnumbered=false,bookmarksopen=true,bookmarksopenlevel=2,
breaklinks=true,pdfborder={0 0 0},pdfborderstyle={},backref=false,colorlinks=true]
{hyperref}
\hypersetup{
	citecolor=blue,linkcolor=blue,urlcolor=blue}



\begin{document}
	
	\title{Quasiclassical Green's function in the path integral approach}
	
	\author{P.A. Krachkov}\email{P.A.Krachkov@inp.nsk.su}
	\affiliation{Budker Institute of Nuclear Physics, SB RAS, Novosibirsk, 630090, Russia}
		
	\date{\today}
\begin{abstract}
We derive the quasiclassical Green's function of a high-energy charged particle in an atomic potential, both alone and in combination with a smooth electromagnetic background, using the path integral approach. We show that the quantum fluctuations essential in the argument of the atomic potential can be replaced by the fluctuation at a single point of closest approach, which allows one to reduce the functional integral to a finite-dimensional one. For a superposition with a smooth background we obtain a general formula in terms of its proper-time kernel and apply it to a plane-wave laser field and to a constant electromagnetic field, in both cases for an arbitrary angle between the particle and the field. Both the scalar and the spinor Green's functions are obtained, the latter by including an additional spin factor along the trajectory. The results are exact in the atomic potential, to all orders in $Z\alpha$, and in the background field parameters, and hold at leading order in $m/\varepsilon$. The obtained Green's functions can be used for the calculation of the cross sections of various QED processes in the combined fields at high energies. The method developed here is of independent interest and can be applied to more complicated systems.	
\end{abstract}
\maketitle
\section{Introduction}\label{sec:intr}
QED processes in the presence of an atomic field play an important role in high-energy physics because they describe the interaction of relativistic particles with matter. This is of critical importance for detector design, especially for calorimeters, and for the planning and interpretation of precision experiments at high-energy facilities. It is required to know the cross sections of these processes to all orders in the atomic field parameter $Z\alpha$ because for heavy atoms, which are often used in high-energy detection systems, $Z\alpha$ is of the order of unity. For ultrarelativistic particles, the main contributions to the cross section usually come from small angles between the momenta of the initial and the final particles, $\theta\sim m/\varepsilon$. The quasiclassical approximation works well in this region. This method allows one to obtain the cross section of a basic process in the field of heavy atoms, with the first corrections in the parameters $\frac{m}{\varepsilon}$ and $\theta$, see the reviews~\cite{KML2016, KM2019}. This powerful method is based on the quasiclassical Green's function $G(\bm{r_{2}},\bm{r_{1}}|\varepsilon)$, which was first obtained for the Coulomb field in Ref.~\cite{MS83} and for an arbitrary atomic potential $V(\bm r)$ in Refs.~\cite{Lee:2000, KM2015, KM2018OperDer}, allowing one to naturally account for effects such as atomic screening or the nuclear form factor:
$$
G(\bm{r_{2}},\bm{r_{1}}|\varepsilon)=\frac{i}{4\pi^{2}r}\int d\bm{q}\,\exp\left[i\kappa r-ir\int_{0}^{1}d u\, V\left(\bm{r}_{1}+\bm{r}u+\bm{q}\sqrt{\frac{2r_{1}r_{2}}{\kappa r}}\right)+iq^{2}\right]\,,
$$
where $\bm{r}=\bm{r}_2-\bm{r}_1$ and $\kappa=\sqrt{\varepsilon^2-m^2}$. The integral over a two-dimensional vector $\bm q$ perpendicular to $\bm r$ is usually interpreted as accounting for quantum fluctuations near the linear classical trajectory. If we neglect the term proportional to $\bm q$ in the argument of the potential, we obtain the eikonal Green's function~\cite{Glauber:1959}. The eikonal approach works well in the case of small gradients. Unfortunately, in the vicinity of an atom, the transverse gradient is not small, which makes this approximation inapplicable.

The path integral formalism is based on the superposition of amplitudes over all possible trajectories. While path integral methods (such as the worldline formalism~\cite{schubert2001perturbative, Dunne2005}) have been highly successful for scattering in smooth background fields (such as intense laser pulses or constant electromagnetic fields), their application to localized atomic potentials remains technically challenging. The $1/r$ singularity prevents a closed-form evaluation of the proper-time kernel, and the standard formalism does not naturally separate the classical trajectory from quantum fluctuations in the high-energy limit.

The difficulties become more severe in the case of a superposition of an atomic potential and an electromagnetic background. Such configurations of fields are relevant for the laser-assisted processes on nuclei or for the electron-atom interaction near a compact astrophysical object possessing strong electromagnetic fields. Both types of QED processes are important: first for planning and performing  experiments in a strong laser field~\cite{Abramowicz2019,yakimenko2019facet,gales2018extreme}; second for the accurate description of QED processes in pulsars and magnetars~\cite{Harding2006,Kaspi2017}.  For the combined field, the Green's function is known for the plane-wave laser field in the specific kinematics of a head-on collision~\cite{DM12,DM14}, which allows one to obtain laser-assisted bremsstrahlung~\cite{KMD19}  and pair production~\cite{DM12,DM14} in this head-on geometry. Realistic configurations involve an arbitrary angle between the particle and the laser propagation direction, where no analogous construction has been available. For a constant electromagnetic field combined with an atomic potential, the situation is even less developed, and we are not aware of an analogous construction.

In this paper, we fill this gap by developing a trajectory-localized fluctuation method within the path integral framework. We demonstrate that essential quantum fluctuations that enter the argument of the atomic potential  are confined to a small spatial region near the scattering center. They can be taken into account with the required accuracy by replacing them with the fluctuation at a single point of closest
approach to the atom. The functional integral then reduces to a finite-dimensional one, with the potential evaluated along the classical trajectory shifted by a single transverse fluctuation. This approach not only provides a clear and simple derivation of the known quasiclassical Green's function for an arbitrary localized potential, but also naturally generalizes to the superposition of atomic and external electromagnetic fields whose proper-time kernel is  known in closed form.

Using this approach, we obtain the general formulas~\eqref{eq:G_general_master} and~\eqref{eq:G_general_master_spin} for the quasiclassical Green's function in an atomic potential combined with a smooth electromagnetic background, together with the conditions of their applicability, and apply them to two backgrounds of physical interest. For a plane-wave laser field, we obtain the Green's function for an arbitrary angle between the particle and the laser propagation direction, Eq.~\eqref{eq:G_laser_x_arb_new}; in the head-on limit it reduces to the known result of Refs.~\cite{DM12,DM14}. For a constant electromagnetic field, we obtain the covariant expression and, for a magnetic field, the Green's function at an arbitrary angle between the particle momentum and the field, including the fixed-energy representation. Since the method applies primarily to the coordinate dependence of the Green's function, we first discuss the scalar particle. We then show that the same framework extends to a spin-$1/2$ particle by introducing an additional spin factor~\eqref{eq:spin_factor1}.

All results are exact in the atomic potential, to all orders in $Z\alpha$, and in the background field parameters, and hold at leading order in $m/\varepsilon$. We assume that the background is smooth on the Compton scale and is less than or of the order of the critical Schwinger value; for the laser field we additionally restrict ourselves to $\xi\lesssim1$, so that the laser-induced transverse momentum remains of order $m$.  The exact Green's functions for a constant electromagnetic field and a plane-wave field are rederived in the path integral formalism in the appendices for completeness.

\section{Quasiclassical Green's function for the scalar particle in an atomic field}\label{sec:QQGF_atom}
We start our consideration from the standard path integral representation of QED Green's function for a relativistic scalar charged particle in an external electromagnetic field described by the four-vector potential $A^{\mu}(x)$, see books~\cite{kleinert2006path,makeenko2002methods}:
\begin{align}\label{eq:Green_start}
	G(x_2,x_1)\simeq\int_{0}^{\infty}ds\,\int\mathcal{D}^{4}y\,\exp\left[-ism^{2}/2-i/2\int_{0}^{s}d\tau\left[y'{}^{2}(\tau)+2y'{}^{\mu}(\tau) e A_{\mu}(y(\tau))\right]\right]\,,
\end{align}
with additional conditions $y(0)=x_1$ and $y(s)=x_2$, where $x^0_{1,2}=t_{1,2}$ and $x^i_{1,2}=r^i_{1,2}$. Although the phase in Eq.~\eqref{eq:Green_start} does not coincide with the standard relativistic Lagrangian $L=-m\sqrt{\dot{y}^2}-e\dot{y}^\mu A_\mu$, it is the polynomial form of the action, obtained in the Fock-Schwinger proper-time representation. This form is equivalent to the classical relativistic action in the sense that it reproduces the same equations of motion and the same value on classical trajectories, while being quadratic in velocities and therefore more convenient for path integral computations. Here we assume $\hbar=c=1$. Throughout this paper, we use the notation $\simeq$ to indicate equality up to pre-exponential factors. Most of them are independent of the external field and can be easily reconstructed from the corresponding free-particle integral. Prefactors with non-trivial dependence on the external field will be discussed separately.

For the Green's function in the atomic field, we have $eA^0(y)=V(\bm y)$ and $A^i(y)=0$. In this case, the integral over $y^0$ is Gaussian and can be simply computed:
\begin{align}
	G(x_2,x_1)\simeq\int_{0}^{\infty}ds\,\int\mathcal{D}^{3}\bm{y}\,
	\exp\left[-i\frac{sm^{2}}{2}+\frac{i}{2}\int_{0}^{s}d\tau\bm{y'^{2}}(\tau)+\frac{i}{2}\int_{0}^{s}d\tau V^{2}(\bm{y}(\tau))-\frac{i}{2s}\left[t+\int_0^s d\tau V(\bm{y}(\tau))\right]^{2}\right]\,,
\end{align}
where $t=t_2-t_1$ and $\bm r=\bm r_{2}-\bm r_{1}$. 

Since in the external time-independent field the energy is conserved, the Green's function at fixed energy $G(\bm{r_2},\bm{r_1}|\varepsilon)$ is more convenient for practical calculations. It can be simply obtained from $G(x_2,x_1)$ by the Fourier transform:
\begin{align}\label{eq:Green_E}
	&G(\bm{r_{2}},\bm{r_{1}}|\varepsilon)=\int dt e^{i\varepsilon t}G(x_2,x_1)\\
	&\simeq\int_{0}^{\infty}ds\,\int\mathcal{D}^{3}\bm{y}
	\exp\left[i\frac{s(\varepsilon^{2}-m^{2})}{2}+\frac{i}{2}\int_{0}^{s}d\tau\bm{y'^{2}}(\tau)+\frac{i}{2}\int_{0}^{s}d\tau V^{2}(\bm{y}(\tau))-i\varepsilon\int_{0}^{s}d\tau V(\bm{y}(\tau))\right]\,.\nonumber
\end{align}
Equation~\eqref{eq:Green_E} is exact  up to field-independent pre-exponential factors. To make it suitable for practical computation, below we introduce several approximations appropriate for the high-energy regime.

Throughout this paper, we assume that the atomic potential $V(\bm r)$ falls off as $1/r$ or faster at large distances and grows no faster than $1/r$ at small distances. The former condition corresponds to the screening of the nuclear charge by atomic electrons at distances beyond the screening radius $r_{scr}$, which in the Thomas--Fermi model is estimated as $r_{scr}\sim 1/(m\alpha Z^{1/3})$, while the latter one corresponds to the finite nuclear size at distances below the nuclear radius $R_{nuc}$. In the intermediate region $R_{nuc}\ll r\ll r_{scr}$ the potential is pure Coulomb, $V(\bm r)=-Z\alpha/r$; for example, the Compton wavelength scale $1/m$ lies well within this Coulomb region.

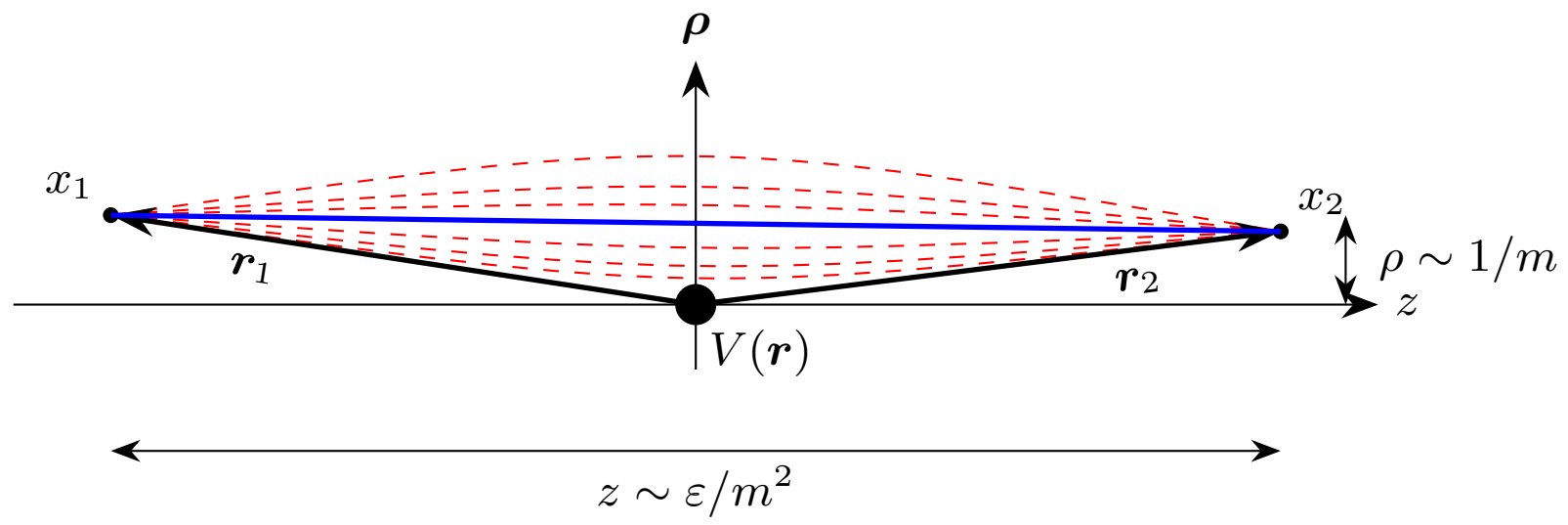
\begin{figure}
	\centering
\begin{tikzpicture}[
	scale=1.1,
	vec/.style={-{Stealth[length=3mm, width=2mm]}, line width=1pt},
	axis/.style={-{Stealth[length=2.5mm]}, thin},
	traj/.style={line width=1pt, blue},
	fluct/.style={dashed, red, thin},
	lbl/.style={font=\small}
	]
	
	% Ось z и поперечное направление
	\draw[axis] (-4.2,0) -- (4.2,0) node[right, lbl] {$z$};
	\draw[axis] (0,-0.4) -- (0,1.5) node[above, lbl] {$\bm\rho$};
	
	% Центр потенциала: маленький закрашенный кружок с подписью
	\filldraw[black] (0,0) circle (0.12);
	\node at (0.4, -0.3) [lbl] {$V(\bm r)$};
	
	% Точки 1 и 2
	\coordinate (p1) at (-3.6, 0.55);
	\coordinate (p2) at (3.6, 0.45);
	
	% КВАНТОВЫЕ ФЛУКТУАЦИИ (фон, рисуем первыми)
	% Ниже классической траектории
	\draw[fluct] (p1) .. controls (0, 0.05) .. (p2);
	\draw[fluct] (p1) .. controls (0, 0.15) .. (p2);
	\draw[fluct] (p1) .. controls (0, 0.3) .. (p2);
	% Выше классической траектории
	\draw[fluct] (p1) .. controls (0, 0.65) .. (p2);
	\draw[fluct] (p1) .. controls (0, 0.8) .. (p2);
	\draw[fluct] (p1) .. controls (0, 1.05) .. (p2);
	
	% Подпись флуктуаций (закомментирована)
	%	\node at (1.4, 1.15) [lbl, red] {fluctuations $\delta\bm\rho$};
	
	% Векторы r1 и r2 от центра (подписи ближе к концам векторов)
	\draw[vec] (0,0) -- (p1) node[pos=0.75, below, sloped, lbl] {$\bm r_1$};
	\draw[vec] (0,0) -- (p2) node[pos=0.75, below, sloped, lbl] {$\bm r_2$};
	
	% Концевые точки
	\filldraw[black] (p1) circle (1.2pt) node[above left, lbl] {$x_1$};
	\filldraw[black] (p2) circle (1.2pt) node[above right, lbl] {$x_2$};
	
	% КЛАССИЧЕСКАЯ ТРАЕКТОРИЯ (поверх флуктуаций, без подписи)
	\draw[traj] (p1) -- (p2);
	
	% Масштаб вдоль z
	\draw[{Stealth[length=2mm]}-{Stealth[length=2mm]}, thin] 
	(-3.6,-0.9) -- (3.6,-0.9);
	\node at (0,-1.15) [lbl] {$z \sim \varepsilon/m^2$};
	
	% Масштаб поперёк
	\draw[{Stealth[length=2mm]}-{Stealth[length=2mm]}, thin] 
	(4.0,0) -- (4.0,0.55);
	\node at (4.1,0.28) [right, lbl] {$\rho\sim 1/m$};
	
\end{tikzpicture}
	\caption{Typical trajectories (red dashed lines) giving the main contribution to the path integral for the high-energy Green's function. The classical trajectory is blue.}
	\label{fig:trajectories}
\end{figure}

In addition to the conditions on the potential, we restrict our consideration to the high-energy kinematic region, where the main contribution to the cross section of QED processes is given by the following region of the Green's function arguments:
\begin{align}\label{eq:kin}
	z,z_{1},z_{2}\sim\frac{\varepsilon}{m^{2}},\quad\rho\sim\frac{1}{m}\,.
\end{align}
Here we set the $\bm z$-axis to be parallel to $\bm r$ and put the origin of the coordinate frame in the center of the atomic potential. In this case, the transverse components of $\bm r_{1,2}$ are the same: $\bm \rho=\bm r_{1\perp}=\bm r_{2\perp}$. We also assume that $z_1 z_2<0$, which corresponds to the fact that the charged particle moves near the potential. Otherwise, the charged particle would move far from the potential, and the eikonal approximation could be used. The typical kinematics of the Green's function arguments is shown in Fig.~\ref{fig:trajectories}. In this paper, we derive the leading-order quasiclassical Green's function, neglecting corrections of order $m/\varepsilon$.

We split the trajectory into the classical part $\bm y_{cl}(\tau)=\bm r_{1}+\bm r \tau/s$ and the quantum fluctuation 
$\delta\bm y(\tau)$: $\bm y(\tau)=\bm y_{cl}(\tau)+\delta\bm y(\tau)$. In this case, the phase in Eq.~\eqref{eq:Green_E} has the following form:
\begin{align}\label{eq:phase}
is\frac{\varepsilon^{2}-m^{2}}{2}+i\frac{r^2}{2s}+\frac{i}{2}\int_{0}^{s}d\tau\delta\bm{y'}^{2}(\tau)-i\varepsilon\int_{0}^{s}d\tau V\left(\bm{y}(\tau)\right)+\frac{i}{2}\int_{0}^{s}d\tau V^{2}\left(\bm{y}(\tau)\right)
\end{align}

Let us estimate each term in the phase. The first two terms are the largest. The characteristic value of the parameter $s$ can be estimated from these two terms using the saddle point approximation: $s\sim\frac{r}{\sqrt{\varepsilon^2-m^2}}\sim\frac{1}{m^2}$. Since $\tau_0 \sim s$ for the high-energy kinematics, we also have $\tau_0 \sim 1/m^2$. The quantum fluctuation term should be of order unity, otherwise the integral will be suppressed, which results in $\delta y\sim 1/m$. Both transverse and longitudinal fluctuations are of the same order, $\delta y\sim 1/m$. However, relative to the corresponding scales of motion, the longitudinal fluctuation is suppressed, $\delta z/z\sim m/\varepsilon\ll 1$, whereas the transverse one is of order unity, $\delta \rho/\rho\sim 1$. Typical trajectories giving the main contribution to the path integral for the high-energy Green's function are shown in Fig.~\ref{fig:trajectories}.

To estimate the last two terms, we can use the classical trajectory in the argument of the potential. The last term can be estimated for the Coulomb case:
\begin{align}\label{eq:estimation_V2}
\int_{0}^{s}d\tau V^{2}(\bm{y}(\tau))\sim \frac{\left(Z\alpha\right)^2 s}{r}\int \frac{dz}{z^2+\rho^2}\sim
\frac{\left(Z\alpha\right)^2m}{\varepsilon}\ll 1\,.
\end{align}
It is small and can be neglected. The other term is logarithmically divergent for the pure Coulomb case. The integral is regularized by the screening radius $r_{scr}$, and the estimation can be written as:
\begin{align}\label{eq:estimation_V}
\varepsilon\int_{0}^{s}d\tau V(\bm{y}(\tau))\sim \frac{Z\alpha \varepsilon s}{r}\int_{-r_{scr}}^{r_{scr}} \frac{dz}{\sqrt{z^2+\rho^2}}\sim
(Z\alpha)\log\left(m r_{scr}\right)\sim 1\,.
\end{align}

Thus, the term linear in $V$ is of order unity and must be kept in the phase, whereas the term quadratic in $V$ is small and will be neglected. 

Let $\tau_0$ be the point where the classical trajectory is closest to the atom, $0=z_1+z\tau_0/s$. We want to show that we can replace $\delta\bm y(\tau)$ by $\delta\bm y(\tau_0)$ in the argument of the atomic potential. It is clear that if $\tau-\tau_0$ is small, the difference of the quantum fluctuations between these two points is also small: $\left(\delta \bm y(\tau)-\delta \bm y(\tau_0)\right)^2\sim|\tau-\tau_0|$, which is related to the fact that the quantum fluctuation term in the phase~\eqref{eq:phase} should be of order unity. For this case, let us estimate this difference:
\begin{align}
&\varepsilon\int d\tau \left(V(\bm{y}_{cl}(\tau)+\delta \bm y(\tau))-V(\bm{y}_{cl}(\tau)+\delta \bm y(\tau_0))\right) =
\varepsilon\int d\tau \left(\delta \bm y(\tau)-\delta \bm y(\tau_0)\right)\cdot\nabla V(\bm{y}_{cl}(\tau)+\delta \bm y(\tau_0)) \nonumber\\
&+\frac{\varepsilon}{2}\int d\tau \left(\delta \bm y(\tau)-\delta \bm y(\tau_0)\right)^i \left(\delta \bm y(\tau)-\delta \bm y(\tau_0)\right)^j \nabla^i\nabla^j V(\bm{y}_{cl}(\tau)+\delta \bm y(\tau_0))\,.
\end{align}
The first integral, which corresponds to the first-order Taylor expansion, is zero after averaging over angles. The second-order Taylor expansion after angular averaging can be estimated as:
\begin{align}\label{eq:V_dif}
	Z\alpha\varepsilon\int\frac{d\tau |\tau-\tau_0|}{\left[(\bm\rho+ \delta \bm y_\perp(\tau_0))^2+(\tau-\tau_0)^2 r^2/s^2\right]^{3/2}}\sim\frac{m}{\varepsilon} \,.
\end{align}
In this integral, the main contribution comes from the region $(\tau-\tau_0)\sim\frac{1}{m\varepsilon}$. 

On the other hand, far from the point $\tau_0$ we can use a Taylor expansion over the fluctuation $\delta\bm y(\tau)$ itself rather than over the difference $\delta\bm y(\tau)-\delta\bm y(\tau_0)$. We introduce the intermediate scale $\lambda\gg\frac{1}{m\varepsilon}$ and consider the region $|\tau-\tau_0|>\lambda$. In this region the classical distance is much larger than the quantum fluctuation $|\bm y_{cl}(\tau)|\approx|z(\tau)|\gg\delta y\sim 1/m$, so we can expand the potential:
\begin{align}
	V(\bm y_{cl}(\tau)+\delta\bm y(\tau))=V(\bm y_{cl}(\tau))
	+\delta\bm y(\tau)\cdot\nabla V(\bm y_{cl}(\tau))
	+\frac{1}{2}\delta y^i(\tau)\delta y^j(\tau)\nabla^i\nabla^j V(\bm y_{cl}(\tau))+\ldots
\end{align}
The linear term vanishes after averaging over angles. The second-order term, integrated over the region $|\tau-\tau_0|>\lambda$, is estimated as:
$$
\frac{\varepsilon}{2}\int\limits_{|\tau-\tau_0|>\lambda}d\tau\,\delta y^2(\tau)\,\nabla^2 V(\bm y_{cl}(\tau))
\sim\frac{Z\alpha}{m^2\varepsilon^2}\int\limits_{|\tau-\tau_0|>\lambda}\frac{d\tau}{|\tau-\tau_0|^3}
\sim\frac{1}{\left(m\varepsilon\lambda\right)^2}\,.
$$
We choose the scale $\lambda=1/\sqrt{\varepsilon m^3}$. In the region $|\tau-\tau_0|>\lambda$, the correction from the quantum fluctuation is small, of order $m/\varepsilon$, as shown above. In the opposite region $|\tau-\tau_0|<\lambda$, the Taylor expansion in the variables $\delta\bm y(\tau)-\delta\bm y(\tau_0)$ works, and the correction is also of order $m/\varepsilon$, see Eq.~\eqref{eq:V_dif}. Thus, in both regions we can simply replace the fluctuation $\delta\bm y(\tau)$ by $\delta\bm y(\tau_0)$ in the argument of the atomic potential.

After this replacement, the path integral over $\delta\bm y(\tau)$ with the point $\tau_0$ separated factorizes into two path integrals, over the intervals $[0,\tau_0]$ and $[\tau_0,s]$. Both are Gaussian and can be simply calculated. The result has the following form:
\begin{align}
	G(\bm{r_{2}},\bm{r_{1}}|\varepsilon)\simeq\int_{0}^{\infty}ds\int d\delta\bm{y}(\tau_{0}) \exp\left[i\frac{s(\varepsilon^{2}-m^{2})}{2}+i\frac{(\bm{r}_{1}-\bm{y}(\tau_{0}))^{2}}{2\tau_{0}}+i\frac{(\bm{r}_{2}-\bm{y}(\tau_{0}))^{2}}{2(s-\tau_{0})}-i\varepsilon\int_{0}^{s}d\tau V(\bm{y}_{cl}(\tau)+\delta \bm y(\tau_0))\right]\,.
\end{align}

 Moreover, we can neglect the quantum fluctuation in the $z$ direction for the following reason. Each term of the Taylor expansion in $\delta y_\parallel(\tau_0)$ is small:
 \begin{align}
 &\frac{\delta y_\parallel(\tau_0)^n}{n!}\varepsilon\int_{0}^{s}d\tau \partial_z^n V(\bm y_{cl}(\tau)+\delta\bm y_\perp(\tau_0))=\frac{s}{z}\frac{\delta y_\parallel(\tau_0)^n}{n!}\varepsilon\int_{z_1}^{z_2}d z \partial_z^n V(\bm\rho+\delta \bm y_\perp(\tau_0), z)=\nonumber\\
 &\frac{s}{z}\frac{\delta y_\parallel(\tau_0)^n}{n!}\varepsilon \partial_z^{n-1} V(\bm\rho+\delta \bm y_\perp(\tau_0), z)|_{z_1}^{z_2}\sim\frac{s\varepsilon}{z}\left(\frac{\delta y_\parallel(\tau_0)}{z}\right)^n\sim\left(\frac{m}{\varepsilon}\right)^n\,.
 \end{align}
 Performing integration over $\delta y_\parallel(\tau_0)$ and assuming that $\bm r_{1\perp}=\bm r_{2\perp}=\bm\rho$, we obtain:
\begin{align}
	&G(\bm{r_{2}},\bm{r_{1}}|\varepsilon)\simeq\int_{0}^{\infty}ds\int d\delta\bm{y}_\perp(\tau_{0})
	\exp\left[i\frac{s\kappa^2}{2}+i\frac{r^{2}}{2s}+i\frac{s\delta\bm{y}^{2}_\perp(\tau_{0})}{2\tau_{0}(s-\tau_0)}-i\varepsilon\int_{0}^{s}d\tau V(\bm{y}_{cl}(\tau)+\delta \bm y_\perp(\tau_0))\right]\,,
\end{align}
where $\kappa=\sqrt{\varepsilon^2-m^2}$. Making the substitution $\delta \bm y_\perp(\tau_0)=\bm q \sqrt{\frac{2r_1 r_2}{\kappa r}}$, changing the variables $\tau =s u$, we obtain:
\begin{align}
G(\bm{r_{2}},\bm{r_{1}}|\varepsilon)=\frac{\sqrt{i}}{2\pi}\int_{0}^{\infty}\frac{ds}{(2\pi s)^{3/2}}\int d\bm{q}
\exp\left[i\frac{s\kappa^2}{2}+i\frac{rq^{2}}{s\kappa}+i\frac{r^{2}}{2s}-i\varepsilon s\int_{0}^{1}du\, V\left(\bm r_1+u\bm r+\bm{q}\sqrt{\frac{2r_{1}r_{2}}{\kappa r}}\right)\right]\,.
\end{align}
Here we restore the prefactor from the free-particle Green's function. This is possible because the integration performed above produces no prefactor depending on the external field. The integration over $s$ gives the result:
\begin{align}
	G(\bm{r_{2}},\bm{r_{1}}|\varepsilon)=\frac{i}{4\pi^{2}}\int \frac{d\bm{q}}{\sqrt{r^2+2q^2r/\kappa}}\,
	\exp\left[i\sqrt{\left[\kappa^2-2\varepsilon\int_{0}^{1}d u\, V\left(\bm{r}_{1}+\bm{r}u+\bm{q}\sqrt{\frac{2r_{1}r_{2}}{\kappa r}}\right)\right]\left[r^{2}+\frac{2rq^{2}}{\kappa}\right]}\right]\,.
\end{align}
Using the estimate of the quantum fluctuations $\delta y_\perp\sim 1/m$ and the substitution $\delta\bm y_\perp(\tau_0)=\bm q\sqrt{2r_1r_2/(\kappa r)}$, we find that the characteristic value of $q$ is of order unity, $q\sim 1$. The result can be simplified: 
\begin{align}\label{eq:G_atom_res}
	G(\bm{r_{2}},\bm{r_{1}}|\varepsilon)=\frac{i}{4\pi^{2}r}\int d\bm{q}\,\exp\left[i\kappa r-ir\int_{0}^{1}d u\, V\left(\bm{r}_{1}+\bm{r}u+\bm{q}\sqrt{\frac{2r_{1}r_{2}}{\kappa r}}\right)+iq^{2}\right]\,.
\end{align}

This formula coincides with the known quasiclassical Green's function obtained in Ref.~\cite{KML2016}. To sum up, the specific high-energy kinematics of the Green's function arguments~\eqref{eq:kin} and the localization of the potential make it possible to reduce quantum fluctuations in the argument of the potential to fluctuations at a single point $\tau=\tau_0$. As a result, the path integral representation of the Green's function reduces to a two-dimensional integral representation. 

This simple physical picture suggests a natural generalization. The single-point fluctuation reduction method developed here can be applied to more general configurations. In particular, it can be used to obtain the Green's function in the superposition of a localized atomic potential and an additional smooth electromagnetic background, provided that the exact Green's function in the background field alone (without the atomic potential) is known. In the next section we derive the corresponding general expression, and then apply it to the laser field and to the constant electromagnetic field.

\section{Green's function in combined atomic and smooth electromagnetic background}\label{sec:general}

In this section, we apply the single-point fluctuation reduction method developed in the previous section to derive the quasiclassical Green's function in the combined atomic and additional smooth electromagnetic background. By ``smooth'' we mean that the background field does not vary significantly on the Compton scale $1/m$. We also assume that the exact Green's function of the scalar particle in this smooth background, $G_0(x_2,x_1)$, is known in the proper-time representation:
\begin{align}\label{eq:G_background}
	G_0(x_2,x_1)=\int_{0}^{\infty}\frac{ds}{2i}\, g_0(x_2,x_1,s)\simeq\int_{0}^{\infty}ds\,\int\mathcal{D}^{4}y\,
	\exp\left[iS_0[y]\right]\,,
\end{align}
i.e. the function $g_0(x_2,x_1,s)$ is known in closed form. Here $S_0$ is the action in the background field alone:
\begin{align}\label{eq:S0}
	S_0[y]=-\frac{sm^2}{2}-\frac{1}{2}\int_{0}^{s}d\tau\,y'^2(\tau)
	-\int_{0}^{s}d\tau\,y'^\mu(\tau)eA_\mu(y(\tau))\,,
\end{align}
where the four-vector potential $A_\mu$ describes the electromagnetic background field.

The prefactor $1/(2i)$ in Eq.~\eqref{eq:G_background} arises from the following fact. We define the Green's function as the solution of $((p-eA)^2-m^2)G(x_2,x_1)=\delta^{(4)}(x_2-x_1)$, which in the proper-time representation reads $1/((p-eA)^2-m^2)=\int_0^\infty \frac{ds}{2i}\,e^{i\frac{s}{2}((p-eA)^2-m^2)}$. This phase convention is consistent with the action definition~\eqref{eq:S0}. The convergence of the integral over $s$ is ensured by the standard prescription $m^2\to m^2-i0$.

The Green's function in the combined field can be written in the following form:
\begin{align}\label{eq:G_combined}
	G(x_2,x_1)\simeq\int_{0}^{\infty}ds\,\int\mathcal{D}^{4}y\,
	\exp\left[iS_0[y]+iS_V[y]\right]\,,
\end{align}
where $S_V$ is the additional phase related to the atomic potential:
\begin{align}
	S_V[y]=-\int_{0}^{s}d\tau\,y'^0(\tau)V(\bm y(\tau))\,.
\end{align}

The classical trajectory $y_{cl}(\tau)$ in the background field satisfies the Lorentz equation:
\begin{align}\label{eq:Lorentz}
	y_{cl}''^\mu(\tau)=eF^{\mu}_{\ \nu}(y_{cl})y_{cl}'^\nu(\tau)\,,\qquad
	F_{\mu\nu}(x)=\partial_\mu A_\nu(x)-\partial_\nu A_\mu(x)\,,
\end{align}
with boundary conditions $y_{cl}(0)=x_1$ and $y_{cl}(s)=x_2$. 
Since this trajectory is not necessarily a straight line, the high-energy kinematics is most naturally formulated in terms of a reference classical trajectory rather than a particular coordinate axis.

We place the origin of the spatial coordinate frame at the center of the atomic potential, so that the potential is simply $V(\bm r)$. Let $\bar x_1$ and $\bar x_2$ be auxiliary endpoints, and let $\bar y_{cl}(\tau)$ be the  classical reference trajectory in the background potential $A^{\mu}$ satisfying $\bar y_{cl}(0)=\bar x_1$ and $\bar y_{cl}(s)=\bar x_2$, which passes exactly through the origin at some instant. The physical arguments of the Green's function are taken in a neighborhood of this reference trajectory: $x^\mu_{1,2}=\bar x^\mu_{1,2}+\Delta x^\mu_{1,2}$ and $\Delta x^\mu_{1,2}\sim\frac{1}{m}$. Then the actual classical trajectory $y_{cl}(\tau)$, connecting $x_1$ and $x_2$, differs from $\bar y_{cl}(\tau)$ by a quantity of order $1/m$. The point of closest approach to the atomic center is characterized by the impact parameter $\bm b=\bm y_{cl}(\tau_0)$, where $\tau_0$ is defined by the condition $\min_\tau |\bm y_{cl}(\tau)| = |\bm y_{cl}(\tau_0)|$, so that $\bm b\cdot\bm y'_{cl}(\tau_0)=0$. 

The quasiclassical regime implies that the arguments of the Green's function satisfy the following conditions:
\begin{align}\label{eq:kin_arb}
	|\bm x_1|,\,|\bm x_2|,\,x^0_1,\,x^0_2 \sim\frac{\varepsilon_{char}}{m^2}\,,\qquad 
	 b\sim\frac{1}{m} \,.
\end{align}
Here $\varepsilon_{char}$ is the characteristic energy of the particle. In the coordinate
representation used in this section the particle does not have a definite energy, so
$\varepsilon_{char}$ is introduced as a characteristic scale for the estimates rather than as an
argument of the Green's function. The quasiclassical estimates give the
relation $\varepsilon_{char}\sim m^2(x^0_2-x^0_1)$ between this scale and the time interval.

We also assume that the typical electromagnetic field is less than or of the order of the critical Schwinger value:
\begin{align}\label{eq:field_strength}
	|eF^{\mu\nu}| \lesssim m^2\,.
\end{align}
This condition ensures that if the initial conditions for the particle satisfy the estimates~\eqref{eq:kin_arb}, the final coordinates will also be of the same order. The condition~\eqref{eq:field_strength} implies that the radius of curvature of the classical trajectory satisfies $R\sim\varepsilon_{char}/|eF|\gtrsim\varepsilon_{char}/m^2$, so that in the formation region $|\tau-\tau_0|\sim 1/(m\varepsilon_{char})$, where the particle moves near the atom at a distance of the order of the Compton wavelength, the trajectory is nearly straight with accuracy $m/\varepsilon_{char}$:
\begin{align}
	\frac{\Delta y'}{y'}\sim\frac{|eF|}{m\varepsilon_{char}}\lesssim\frac{m}{\varepsilon_{char}}\ll 1\,.
\end{align}
Thus, in the formation region near $\tau_0$, the distance to the nucleus grows as follows:
\begin{align}\label{eq:distance_growth}
	|\bm y_{cl}(\tau)|^2\simeq b^2+|{\bm y}'_{cl}(\tau_0)|^2(\tau-\tau_0)^2\,.	
\end{align}
We also require that $|\bm y'_{cl}(\tau_0)|\sim\varepsilon_{char}$.

In addition, the trajectory should not return to the atomic region after the closest approach:
\begin{align}\label{eq:no_return}
	|{\bm y}_{cl}(\tau)|\gtrsim\frac{1}{m}
	\quad\text{for}\quad 
	|\tau-\tau_0|\gg\frac{1}{m\varepsilon_{char}}\,.
\end{align}
Condition~\eqref{eq:no_return} excludes closed trajectories, such as a purely transverse circular orbit in a magnetic field.

We decompose the path into the classical trajectory and quantum fluctuations:
\begin{align}
	y^\mu(\tau)=y_{cl}^\mu(\tau)+\delta y^\mu(\tau)\,,
\end{align}
where $\delta y(0)=\delta y(s)=0$. 

The estimates from the previous section remain the same:
\begin{align*}
	s\sim\tau_0\sim\frac{1}{m^2}\,,\quad\delta y\sim\frac{1}{m}\,,\quad|\delta y(\tau)-\delta y(\tau_0)| \sim|\tau-\tau_0|^{1/2}\,,
\end{align*}
The phase $S_V[y]$ can be estimated by analogy with the previous section: $S_V[y]\sim S_V[y_{cl}]\sim1$. Thus, we can replace ${y'}^0(\tau)\to {y'}_{cl}^0(\tau)$ in $S_V[y]$, because ${y'}_{cl}^0(\tau)\sim\varepsilon_{char}\gg{\delta y'}^0(\tau)\sim m$.

We want to show that we can replace $\delta\bm y(\tau)$ by $\delta\bm y(\tau_0)$ in the argument of the atomic potential. For brevity, we denote $\delta y_*^{\mu} \equiv \delta y^{\mu}(\tau_0)$ and $ y_*^{\mu} \equiv  y^{\mu}(\tau_0)$. The proof of this statement is analogous to the case of the pure atomic potential, presented in the previous section. We recommend that readers first review the analogous proof in the previous section, because here we will not repeat all the details.

We consider the difference:
\begin{align}\label{eq:red_fluct}
	&\int d\tau\, {y'}_{cl}^0(\tau)
	\left[
	V(\bm y_{cl}(\tau)+\delta\bm y(\tau))
	-V(\bm y_{cl}(\tau)+\delta\bm y_*)
	\right]\nonumber\\
	&=\int d\tau\, {y'}_{cl}^0(\tau)
	\left(\delta\bm y(\tau)-\delta\bm y_*\right)\cdot
	\nabla V(\bm y_{cl}(\tau)+\delta\bm y_*)\nonumber\\
	&\quad+\frac{1}{2}\int d\tau\,{y'}_{cl}^0(\tau)
	\left(\delta\bm y(\tau)-\delta\bm y_*\right)^i
	\left(\delta\bm y(\tau)-\delta\bm y_*\right)^j
	\nabla^i\nabla^j V(\bm y_{cl}(\tau)+\delta\bm y_*)\,.
\end{align}
The first term vanishes after angular averaging. In order to estimate the second term we split the integral over $\tau$ into two regions:  the first region is $|\tau-\tau_0|<\lambda$, and the second region is $|\tau-\tau_0|>\lambda$, where $\lambda=1/\sqrt{\varepsilon_{char} m^3}$. 

In the first region, we use $\delta\bm y(\tau)-\delta\bm y_*\sim|\tau-\tau_0|^{1/2}$ and apply the condition~\eqref{eq:distance_growth}. The result for the second term of~\eqref{eq:red_fluct} is estimated as:
\begin{align}\label{eq:V_dif_const}
	Z\alpha\int
	\frac{d\tau\,|\tau-\tau_0| {y'}_{cl}^0(\tau)}
	{\left[(\bm{b}+\delta \bm{ y}_{*\perp})^2+(\bm y'_{cl}(\tau_0)(\tau-\tau_0)+\delta \bm{ y}_{*\parallel})^2\right]^{3/2}}
	\sim\frac{m}{\varepsilon_{char}}\,.
\end{align}
The main contribution to this integral comes from the region $|\tau-\tau_0|\sim 1/(m\varepsilon_{char})$.

In the second region, $|\bm y_{cl}(\tau)|\gg|\delta\bm y_*|$, we can omit $\delta\bm y_*$ in the argument of the potential. In this case, the expression can be estimated as follows:
\begin{align}
	&
	\int\limits_{|\tau-\tau_0|>\lambda}d\tau\,{y'}_{cl}^0(\tau)
	\delta y^2(\tau)\,\nabla^2 V(\bm y_{cl}(\tau))
	\sim
	\frac{Z\alpha}{m^2}
	\int\limits_{|\tau-\tau_0|>\lambda}d\tau\,\frac{{y'}_{cl}^0(\tau)
	}{|\bm y_{cl}(\tau)|^3}
	\sim \frac{Z\alpha}{2 m^2 |\bm y_{cl}(\tau_0+\lambda)|^2}\sim
	\frac{m}{\varepsilon_{char}}\,.
\end{align}

In both regions the correction is of order $m/\varepsilon_{char}$, so in the leading quasiclassical approximation the replacement 
$\delta\bm y(\tau)\to\delta\bm y(\tau_0)\equiv\delta\bm y_*$ in the argument of the atomic potential is valid.

Let $\bm \nu(\tau)$ be the unit vector parallel to $\frac{d \bm y_{cl}(\tau)}{d\tau}$. For
an arbitrary vector $\bm x$ we set $\bm x_\parallel(\tau)=\bm \nu(\tau)
(\bm \nu(\tau)\cdot \bm x)$ and $\bm x_\perp(\tau)=\bm x-\bm \nu(\tau)
(\bm \nu(\tau)\cdot \bm x)$, where the dependence on $\tau$ comes from the direction
$\bm \nu(\tau)$. In particular, $\delta\bm y_{*\parallel}(\tau)$ depends on $\tau$ although
$\delta\bm y_*$ itself is fixed at $\tau_0$.

Consider the first term of the Taylor expansion in $\delta\bm y_{*\parallel}(\tau)$:
\begin{align}
	&\int_{0}^{s}d\tau\,{y'}_{cl}^0(\tau) \delta\bm y_{*\parallel}(\tau)
	\frac{d }{d\bm \nu(\tau)}V(\bm y_{cl}(\tau)+\delta\bm y_{*\perp})
	=\int_{0}^{s}d\tau\,\frac{\delta y_{*\parallel}(\tau) {y'}_{cl}^0(\tau)}{{y'}_{cl\parallel}(\tau)}
\frac{d }{d\tau}V(\bm y_{cl}(\tau)+\delta\bm  y_{*\perp})\nonumber\\
	&=\frac{\delta y_{*\parallel}(\tau){y'}_{cl}^0(\tau)}{{y'}_{cl\parallel}(\tau)}
V(\bm y_{cl}(\tau)+\delta\bm  y_{*\perp}){\Large|}_0^s-
\int_{0}^{s}d\tau\, V(\bm y_{cl}(\tau)+\delta\bm  y_{*\perp})
\frac{d }{d\tau}\frac{\delta y_{*\parallel}(\tau){y'}_{cl}^0(\tau)}{{y'}_{cl\parallel}(\tau)}\,.
\end{align}
The first term in the last line is of the order of $\frac{m}{\varepsilon_{char}}$ for the Coulomb potential and less for the screening potential. To estimate the second term, we use the equation of motion~\eqref{eq:Lorentz} together with the field strength condition~\eqref{eq:field_strength}. This allows us to estimate $\frac{d }{d\tau}\frac{\delta y_{*\parallel}(\tau){y'}_{cl}^0(\tau)}{{y'}_{cl\parallel}(\tau)}\sim\frac{|eF^{\mu\nu}|}{m}\lesssim m$. Thus, the second integral is also of the order of $\frac{m}{\varepsilon_{char}}$. The $n$-th term of the Taylor expansion in $\delta\bm y_{*\parallel}(\tau)$ can be estimated in the same way; it is of the order of $\left(\frac{m}{\varepsilon_{char}}\right)^n$. Therefore, we can neglect $\delta\bm y_{*\parallel}(\tau)$ in the argument of the atomic potential. 

The contribution of quantum fluctuations to the atomic potential is localized  in the formation region $|\tau-\tau_0|\sim 1/(m\varepsilon_{char})$, where the classical trajectory curvature is negligible. Therefore, we can define the transverse 
fluctuation  in the argument of the potential  relative to the fixed direction $\bm \nu(\tau_0)$.

To sum up, the phase $S_V$ in the Green's function~\eqref{eq:G_combined} can be significantly simplified:
\begin{align}
	S_V=
	-\int_{0}^{s}d\tau\,{y'}_{cl}^0(\tau)
	V(\bm y_{cl}(\tau)+\delta\bm  y_{*\perp})\,,
\end{align}
with accuracy $m/\varepsilon_{char}$.

The path integral then factorizes into two path integrals corresponding to the intervals $[0,\tau_0]$ and $[\tau_0,s]$, with fixed boundary conditions $y^{\mu}(\tau_0)=y_*^{\mu}$. The result takes the form:
\begin{align}\label{eq:G_general_master}
	&G(x_2,x_1)=
	\int_{0}^{\infty}\frac{ds}{2i}
	\int d^{4}y_*\,
	g_0\!\left(x_2,y_*;s-\tau_0\right)
	g_0\!\left(y_*,x_1;\tau_0\right)\nonumber\\
	&\times\exp\left[-i\int_{0}^{s}d\tau\,{y'}_{cl}^0(\tau)\,
	V(\bm y_{cl}(\tau)+\delta\bm  y_{*\perp})\right]\,,
\end{align}
where $g_0(x_2,x_1;s)$ denotes the proper-time integrand of the background Green's function, Eq.~\eqref{eq:G_background}. Here $\tau_0$ is the proper time of closest approach to the atomic center, $\min_\tau |\bm y_{cl}(\tau)| = |\bm y_{cl}(\tau_0)|$, where the classical trajectory $y_{cl}(\tau)$ is determined by the Lorentz equation with boundary conditions $y_{cl}(0)=x_1$ and $y_{cl}(s)=x_2$. Thus, $\tau_0$ depends on $s$ and the endpoints $x_{1,2}$, i.e., $\tau_0=\tau_0(s;x_1,x_2)$. Note that the overall prefactor and measure normalization will be consistently fixed by requiring that Eq.~\eqref{eq:G_general_master} reduces exactly to $G_0(x_2,x_1)$ when $V=0$, via the composition property of the proper-time kernel.

The transverse fluctuation in the argument of the atomic potential does not have to be defined with respect to exactly $\bm\nu(\tau_0)$. Let $\bm \nu_0$ be any fixed unit vector such that the angle between $\bm \nu_0$ and $\bm\nu(\tau_0)$ is $\theta\lesssim m/\varepsilon_{char}$. The projection $\delta\bm y_{*\perp}$ defined relative to $\bm \nu_0$ then differs from the projection relative to $\bm\nu(\tau_0)$ by a quantity of order $|\delta \bm y_{*\perp}|\,\theta\sim 1/\varepsilon_{char}$, which is small compared to $|\delta \bm y_{*\perp}|\sim1/m$ and produces only a correction of order $m/\varepsilon_{char}$ in the phase. Therefore, within the accuracy of the present approximation, the transverse plane may be chosen in the most convenient way, provided it is almost perpendicular to the classical trajectory at the point of closest approach.

To conclude, we have reduced the functional integral to a single four-dimensional integral over the intermediate point $y_*$, with the atomic potential evaluated along the classical trajectory in the background field. The resulting expression~\eqref{eq:G_general_master} provides a general tool for computing the quasiclassical Green's function in the combined field, provided that the proper-time kernel $g_0$ of the background field alone is known in closed form. This applies, in particular, to such configurations as a plane-wave laser field, a constant electromagnetic field, and their combinations. In the next sections, we apply this formula to the cases of a laser field background and a constant electromagnetic field background.

\section{Green's function in combined atomic and plane-wave laser field}\label{sec:laser_atomic}
 
The laser field propagates along the  unit vector $\bm n$.  For an arbitrary vector $\bm x$, we define the transverse projection relative to the laser propagation direction as $\bm x_{n\perp}=\bm x-\bm n(\bm n\cdot\bm x)$.  The plane-wave vector potential is $\bm A(t-\bm n\cdot\bm r)$, with the condition $\bm n\cdot\bm A=0$.  It is convenient to use the light-cone variables associated with the laser propagation
direction $\bm n$,
\begin{align}\label{eq:lightcone_main}
	\phi=t-\bm n\cdot\bm r\,,\qquad T=\frac{t+\bm n\cdot\bm r}{2}\,,\qquad
	\bm r_{n\perp}=\bm r-\bm n(\bm n\cdot\bm r)\,,
\end{align}
so that $\bm A(t-\bm n\cdot\bm r)=\bm A(\phi)$ and  $y^2=2\phi T-\bm y_{n\perp}^2$, see Appendix~\ref{app:laser}. We denote by $\phi_{1,2}$, $T_{1,2}$, $\bm x_{1,2n\perp}$ the light-cone components of the endpoints $x_{1,2}$.

For a plane wave the field-strength condition~\eqref{eq:field_strength} constrains the combination $\xi\omega_0/m=E_0/E_{\rm cr}\lesssim1$, where $\xi=|e|E_0/(m\omega_0)$ is the dimensionless parameter characterizing the laser field strength, with $E_0$ and $\omega_0$ being the laser electric field amplitude and angular frequency. In what follows we additionally assume $\xi\lesssim1$.  The more general case $\xi\gg1$ can, in principle, be treated within the same general formula~\eqref{eq:G_general_master}, but the quiver motion then raises the transverse momentum scale to $\xi m$ and introduces the effective mass $m_*\sim m\xi$, which modifies both the formation length and the geometry of the closest approach. Such a generalization requires a separate careful analysis and is beyond the scope of the present paper.

The computation of the Green's function in the pure laser plane-wave field is presented in Appendix~\ref{app:laser}. Substituting the exact Green's function in the laser field~\eqref{eq:G_laser_x_app} into the general expression~\eqref{eq:G_general_master}, we obtain the quasiclassical Green's function in the combined laser and atomic field. The action $S_0$ in the plane-wave background satisfies the identity $S_0[y_{cl}+\delta y]=S_0[y_{cl}]+\delta S[\delta y]$ and $S_0[y_{cl}](s,\tau_0)+S_0[y_{cl}](\tau_0,0)=S_0[y_{cl}](s,0)$, where $y_{cl}$ is the classical trajectory, which corresponds to the extremum of the action and $S_0[y_{cl}](a,b)$ denotes the action along $y_{cl}$ on the proper-time interval $[b,a]$. Thus, passing from the integration over $y_*$ to the integration over the fluctuations $\delta y_*$ at this point leads to a significant simplification of the result:
\begin{align}\label{eq:G_laser_x_arb}
	&G(x_2,x_1)=\frac{i}{2}\int_{0}^{\infty}\frac{ds}{(2\pi)^4\tau_0^2(s-\tau_0)^2}
	\int d^4\delta y_*\,\\
	&\times\exp\left[iS_0(x_2,x_1;s)
	-i\int_{0}^{s}d\tau\,{y'}_{cl}^{0}(\tau)\,
	V(\bm y_{cl}(\tau)+\delta\bm y_{*\perp})
	-\frac{i s\,\delta y_*^2}{2\tau_0(s-\tau_0)}
	\right]\nonumber\,,
\end{align}
where $S_0$ is the classical action in the plane-wave~\eqref{eq:G_laser_x_app}. The fluctuation $\delta\bm y_{*\perp}$ in the atomic potential is the transverse component relative to the particle trajectory direction at the point of closest approach, as defined in Section~\ref{sec:general}. 
For an arbitrary angle between the particle momentum and $\bm n$, the transverse plane relative to $\bm n$ generally differs from the transverse plane relative to the particle trajectory.

Since the atomic potential depends only on the transverse fluctuation, the integration over $\delta y_*^0$ and $\delta y_{*\parallel}$ is Gaussian and gives
\begin{align}\label{eq:G_laser_x_arb_new}
	&G(x_2,x_1)=\frac{i}{2\pi}\int_{0}^{\infty}\frac{ds}{(2\pi s)^2}
	\int d^2 q\,
	\exp\left[iS_0(x_2,x_1;s)
	-i\int_{0}^{s}d\tau\,{y'}_{cl}^{0}(\tau)\,
	V\left(\bm y_{cl}(\tau)+\sqrt{\frac{2\tau_0(s-\tau_0)}{s}}\,\bm q\right)
	+iq^2
	\right]\,,
\end{align}
where $\bm q$ is a two-dimensional vector transverse to the spatial velocity $\bm y'_{cl}(\tau_0)$.

In the special case where the particle almost counterpropagates to the laser field direction, which corresponds to the angle between $-\bm n$ and $\bm x_2-\bm x_1$ being of order $m/\varepsilon_{char}$, the result is significantly simplified. We introduce the $z$-axis along $-\bm n$, so that $\bm e_z=-\bm n$. In this geometry, the transverse plane relative to the particle trajectory coincides with the transverse plane relative to $\bm n$. Consequently, $T=\frac{t-z}{2}\sim\frac{1}{\varepsilon_{char}}$ and $\phi=t+z\sim 2z$. The atomic potential $V(\bm y_{cl}(\tau)+\delta\bm y_{*\perp})\simeq V(\phi_{cl}(\tau)/2,\bm y_{cl\perp}(\tau)+\delta\bm y_{*\perp})$ does not depend on $T$.

The classical trajectory and the argument of the atomic potential can be written explicitly. Using $\frac{\tau_0}{s}\simeq\frac{|\phi_1|}{\phi_2-\phi_1}$ and 	${y'}^0_{cl}(\tau)\simeq\frac{\phi_2-\phi_1}{2s}$, Eq.~\eqref{eq:G_laser_x_arb_new} takes the form:
\begin{align}\label{eq:G_laser_atom5}
	&G(x_2,x_1)=\frac{i}{2\pi }\int_{0}^{\infty}\frac{ds}{(2\pi s)^2}\,\int d^{2} q 
	\exp\left[i q^2 -i\frac{(\phi_2-\phi_1)}{2}\int_{0}^{1}d u V(\phi_u/2,\bm y_{u\perp} )\right]\,\\
	\times&\exp\left[-i\frac{sm^{2}}{2}-i\frac{(\phi_2-\phi_1)(T_2-T_1)}{s}
	+\frac{i}{2s}\left[\bm x_{2\perp}-\bm x_{1\perp}+s\int_{0}^{1}du\,e \bm  A(\phi_{u})\right]^{2}
	-i\frac{s}{2}\int_{0}^{1}du\,e^2\bm A^{2}(\phi_{u})
	\right]\,,\nonumber		
\end{align}
where $\bm y_{u\perp}=\bm x_{1\perp} + u\left[\bm x_{2\perp}-\bm x_{1\perp}+s\int_{0}^{1}du'\,e \bm  A(\phi_{u'})\right] - s\int_{0}^{u}du'\,e \bm  A(\phi_{u'})+\sqrt{2 s |\phi_1\phi_2|/(\phi_2-\phi_1)^2}\,\bm q$ and $\phi_u=\phi_1+u(\phi_2-\phi_1)$. Here $\bm q$ is the dimensionless vector characterizing the transverse quantum fluctuations  $\bm{\delta y}_{*\perp}$, and $\bm y_{u\perp}$ is the effective classical  trajectory in the laser field shifted by the quantum fluctuation. We also assume $\phi_1<0<\phi_2$, so that the particle passes near the atom.

The Green's function at fixed light-cone energy  $\varepsilon_{T}$ is obtained by the Fourier transform over $\Delta T=T_2-T_1$:
\begin{align}\label{eq:G_laser_atom_energy}
	&G(\bm x_{2\perp},\phi_{2};\bm x_{1\perp},\phi_{1}|\varepsilon_{T})
	=\frac{i}{4\pi^2(\phi_{2}-\phi_{1})}\int d^{2} q 
	\exp\left[i q^2 -i\frac{(\phi_2-\phi_1)}{2}\int_{0}^{1}d u V(\phi_u/2,\bm \tilde y_{u\perp} )\right]\,\\
	&\times\exp\left[-i\frac{\phi_{2}-\phi_{1}}{2\varepsilon_{T}}\left(m^2
	-\left[\varepsilon_{T} \frac{\bm x_{2\perp}-\bm x_{1\perp}}{\phi_{2}-\phi_{1}}+\int_{0}^{1}du\,e \bm  A(\phi_{u})\right]^{2}
	+\int_{0}^{1}du\,e^2\bm A^{2}(\phi_{u})
	\right)\right]\,,\nonumber
\end{align}
where $\bm \tilde y_{u\perp}=\bm x_{1\perp} + u\left[\bm x_{2\perp}-\bm x_{1\perp}+\frac{\phi_2-\phi_1}{\varepsilon_T}\int_{0}^{1}du'\,e\bm  A(\phi_{u'})\right] - \dfrac{\phi_2-\phi_1}{\varepsilon_T}\int_{0}^{u}du'\, e\bm A(\phi_{u'})+ \sqrt{\dfrac{2|\phi_1\phi_2|}{\varepsilon_T(\phi_2-\phi_1)} }\,\bm q$.

The quasiclassical Green's function at fixed light-cone energy  $\varepsilon_{T}$ for counterpropagation kinematics~\eqref{eq:G_laser_atom_energy} coincides with the analogous result obtained in~\cite{DM12,DM14}, whereas Eq.~\eqref{eq:G_laser_x_arb_new} generalizes it to an arbitrary angle between the particle and laser propagation directions. Note that if we set $V(\bm r)=0$, the result~\eqref{eq:G_laser_atom_energy} coincides with the exact Green's function in the laser field~\eqref{eq:G_laser_final_app}. Moreover, using the following relation valid to leading order in $\frac{m}{\varepsilon}$:
\begin{align}\label{eq:id}
	&G(\bm{r_{2}},\bm{r_{1}}|\varepsilon)=\int d(t_2-t_1) \mbox{e}^{i \varepsilon (t_2-t_1)} G(x_2,x_1)=\int d(t_2-t_1) \mbox{e}^{i \varepsilon (t_2-t_1)}\nonumber\\ &\times\int\frac{d\varepsilon_T}{2\pi}\mbox{e}^{-i\varepsilon_T(T_2-T_1)} 
	G(\bm x_{2\perp},\phi_{2};\bm x_{1\perp},\phi_{1}|\varepsilon_T)=2\mbox{e}^{i\varepsilon(z_2-z_1)}G(\bm x_{2\perp},\phi_{2};\bm x_{1\perp},\phi_{1}|2\,\varepsilon)\,,
\end{align}
the result is in agreement with the quasiclassical Green's function in the atomic field~\eqref{eq:G_atom_res}.

\section{Green's function in combined atomic and constant electromagnetic fields}\label{sec:const_atomic}

In this section, we consider the quasiclassical Green's function of a scalar particle in a superposition of an atomic field, described by the potential $V(\bm r)$, and a constant homogeneous electromagnetic field, described by the four-vector potential $A_\mu(x)=-\frac{1}{2}F_{\mu\nu}x^\nu$. 

The computation of the Green's function in a pure constant homogeneous electromagnetic field is presented in Appendix~\ref{app:const}. Substituting the exact Green's function in the constant field~\eqref{eq:G_const_app} into the general expression~\eqref{eq:G_general_master}, we obtain the quasiclassical Green's function in the combined fields. Since the action in the constant homogeneous electromagnetic field is quadratic, it is convenient to pass to the fluctuations at the point $\tau_0$, $y_*=y_{cl}(\tau_0)+\delta y_*$, see Eq.~\eqref{eq:G_general_master}.  Due to the stationarity of the action at $y_{cl}(\tau_0)$, all terms linear in $\delta y_*$ cancel, and the sum of the two actions reduces to the action along the whole classical trajectory from $x_1$ to $x_2$ plus a quadratic form in $\delta y_*$.  This leads to a significant simplification of the result:
\begin{align}\label{eq:G_const_x_fluct_cov}
	&G(x_2,x_1)=\frac{i}{2}\,e^{i\Phi(x_2,x_1)}
	\int_{0}^{\infty}\frac{ds\mathcal{D}(\tau_0)\,\mathcal{D}(s-\tau_0)}{(2\pi)^4\tau_0^2(s-\tau_0)^2}\int d^4\delta y_*\,
	\\
	&\times\exp\left[-i\frac{sm^{2}}{2}
	-\frac{i}{4}\,x_-^\mu\Big[eF\coth\!\big(eFs/2\big)\Big]_{\mu\nu}x_-^\nu
	-i\int_{0}^{s}d\tau{y'}_{cl}^{0}(\tau) V(\bm y_{cl}(\tau)+\delta\bm y_{*\perp})
	\right]\nonumber
	\\
	&\times\exp\left[
	-\frac{i}{4}\,\delta y_*^\mu
	\left[eF\,\frac{\sinh\!\big(eFs/2\big)}{\sinh\!\big(eF\tau_0/2\big)\sinh\!\big(eF(s-\tau_0)/2\big)}\right]_{\mu\nu}
	\delta y_*^\nu
	\right]\,,\nonumber
\end{align}	
where $\Phi$ is the gauge-dependent Schwinger phase~\eqref{eq:Schwinger_phase}, $\mathcal{D}(\tau)$ is a field-dependent functional determinant~\eqref{eq:det_const_explicit}, and $x_-=x_2-x_1$. Note that we cannot use the representation~\eqref{eq:G_const_app_EB}, which is written in the special reference frame where $\bm E \parallel \bm B$: the atomic potential $V(\bm r)$ is defined in the rest frame of the atom, and a Lorentz boost would change its form. This is why we start from the covariant expression~\eqref{eq:G_const_app}.

In the general case, the Gaussian integration over $\delta y_*^0$ and $\delta y_{*\parallel}$ can be performed explicitly, where $\delta y_{*\parallel}=\delta \bm y_*\cdot \bm y'_{cl}(\tau_0)/|\bm y'_{cl}(\tau_0)|$ is the component of the fluctuation parallel to the classical velocity at the point $\tau_0$. However, the result is cumbersome, and we do not present it here.

We consider the special case where $\bm E \parallel \bm B$ and the angle $\theta$ between $\bm x_2-\bm x_1$ and $\bm B$ is small, $\theta\sim m/\varepsilon_{char}$. In this case we can set $\bm e_z=\frac{\bm B}{B}$, and assume that $\delta\bm y_{*\perp}$ is perpendicular to $\bm e_z$.

\begin{align}\label{eq:G_B_E_parallel_start}
	&G(x_2,x_1)=\frac{i}{2}\,e^{i\Phi(x_2,x_1)}
	\int_{0}^{\infty}\frac{ds\mathcal{D}(\tau_0)\,\mathcal{D}(s-\tau_0)}{(2\pi)^4\tau_0^2(s-\tau_0)^2}\int d^4\delta y_{*} \,\exp\left[-i\frac{sm^{2}}{2}
	-i\int_{0}^{s}d\tau{y'}_{cl}^{0}(\tau) V(\bm y_{cl}(\tau)+\delta\bm y_{*\perp})\right]
	\\
	\times&\exp\left[-i\frac{e\mathcal{E}}{4}\coth\!\left(\frac{e\mathcal{E}s}{2}\right)((x^0_{-})^{2}-(x^3_{-})^{2})
	+i\frac{e\mathcal{B}}{4}\cot\!\left(\frac{e\mathcal{B}s}{2}\right)\bm x_{-\perp}^{2}
	\right]\nonumber	
	\\
	\times&\exp\left[-i\frac{e\mathcal{E}}{4}\frac{\sinh\!\left(\frac{e\mathcal{E}s}{2}\right)}{\sinh\!\left(\frac{e\mathcal{E}\tau_0}{2}\right)\sinh\!\left(\frac{e\mathcal{E}(s-\tau_0)}{2}\right)}
	((\delta y^0_{*})^{2}-(\delta y^3_{*})^{2})
	+i\frac{e\mathcal{B}}{4}
	\frac{\sin\!\left(\frac{e\mathcal{B}s}{2}\right)}{\sin\!\left(\frac{e\mathcal{B}\tau_0}{2}\right)\sin\!\left(\frac{e\mathcal{B}(s-\tau_0)}{2}\right)}
	\delta y_{*\perp}^{2}
	\right]\,,\nonumber	
\end{align}
 where $\mathcal{E}$, $\mathcal{B}$ are field invariants defined in~\eqref{eq:eff_fields}.

Integrating over $\delta y^{0}_{*}$, $\delta y^{3}_{*}$ and introducing the new variable $\bm q=\delta \bm y_{*\perp}/c(s)$, we obtain:
\begin{align}\label{eq:G_B_E_parallel}
	&G(x_2,x_1)=i\,e^{i\Phi(x_2,x_1)}
	\int_{0}^{\infty}\frac{ds}{(2\pi)^3s^2}\int d \bm q \,\exp\left[-i\frac{sm^{2}}{2}
	-i\int_{0}^{s}d\tau{y'}_{cl}^{0}(\tau) V(\bm y_{cl}(\tau)+c(s) \bm q)\right]
	\\
	&\frac{e\mathcal{E}s/2}{\sinh(e\mathcal{E}s/2)}\,
	\frac{e\mathcal{B}s/2}{\sin(e\mathcal{B}s/2)}\,
	\exp\left[-i\frac{e\mathcal{E}}{4}\coth\!\left(\frac{e\mathcal{E}s}{2}\right)((x^0_{-})^{2}-(x^3_{-})^{2})
	+i\frac{e\mathcal{B}}{4}\cot\!\left(\frac{e\mathcal{B}s}{2}\right)\bm x_{-\perp}^{2}+i\bm q^2
	\right]\,,\nonumber	
\end{align}
where $c(s)=\sqrt{\frac{4}{e\mathcal{B}}	\frac{\sin\!\left(\frac{e\mathcal{B}\tau_0}{2}\right) \sin\!\left(\frac{e\mathcal{B}(s-\tau_0)}{2}\right)}{\sin\!\left(\frac{e\mathcal{B}s}{2}\right)}}$.

For  $\bm E=0$ and a small angle $\theta\sim m/\varepsilon_{char}$ between $\bm B$ and $\bm x_2-\bm x_1$, the Green's function is significantly simplified, because the energy is conserved. The time component and the component parallel to $\bm B$ (parallel to the $z$ axis)  of the classical trajectory are $y^{0,3}_{cl}(\tau)=x_1^{0,3}+\frac{\tau}{s}(x_2^{0,3}-x_1^{0,3})$, and 
\begin{align}\label{eq:ycl_perp}
	\bm y_{cl,\perp}(\tau) &= \bm x_{1\perp}+
	\frac{1}{\sin\!\left(\frac{e\mathcal{B}s}{2}\right)}
	\begin{pmatrix}
		\cos\!\left(\frac{e\mathcal{B}(s-\tau)}{2}\right)\sin\!\left(\frac{e\mathcal{B}\tau}{2}\right) & -\sin\!\left(\frac{e\mathcal{B}(s-\tau)}{2}\right)\sin\!\left(\frac{e\mathcal{B}\tau}{2}\right) \\
		\sin\!\left(\frac{e\mathcal{B}(s-\tau)}{2}\right)\sin\!\left(\frac{e\mathcal{B}\tau}{2}\right) & \cos\!\left(\frac{e\mathcal{B}(s-\tau)}{2}\right)\sin\!\left(\frac{e\mathcal{B}\tau}{2}\right)
	\end{pmatrix}
	(\bm x_{2\perp}-\bm x_{1\perp})\,.
\end{align}
The Green's function has the following form:	
\begin{align}\label{eq:G_B_t}
	&G(x_2,x_1)=i\,e^{i\Phi(x_{2\perp},x_{1\perp})}
	\int_{0}^{\infty}\frac{ds}{(2\pi)^3s^2}\int d \bm q \,\exp\left[-i\frac{sm^{2}}{2}
	-i\frac{x_2^0-x_1^0}{s}\int_{0}^{s}d\tau V(\bm y_{cl}(\tau)+c(s) \bm q)\right]
	\\
	&\frac{e\mathcal{B}s/2}{\sin(e\mathcal{B}s/2)}\,
	\exp\left[-i\frac{((x^0_{-})^{2}-(x^3_{-})^{2})}{2s}
	+i\frac{e\mathcal{B}}{4}\cot\!\left(\frac{e\mathcal{B}s}{2}\right)\bm x_{-\perp}^{2}
	+i\bm q^2
	\right]\,.\nonumber	
\end{align}

This Green's function depends only on $x_2^{0}-x_1^{0}$. We can pass to the Green's function at fixed energy $\varepsilon$: 	
\begin{align}\label{eq:G_energy_B_s}
	&G(\bm x_2,\bm x_1|\varepsilon)=\frac{e^{i\frac{\pi}{4}}}{2\pi}\,e^{i\Phi(x_{2\perp},x_{1\perp})}
	\int_{0}^{\infty}\frac{ds}{(2\pi s)^{3/2}}\int d \bm q \,\exp\left[i\frac{s(\varepsilon^2-m^{2})}{2}
	-i\varepsilon\int_{0}^{s}d\tau V(\bm y_{cl}(\tau)+c(s) \bm q)\right]
	\\
	&\frac{e\mathcal{B}s/2}{\sin(e\mathcal{B}s/2)}\,
	\exp\left[+i\frac{(x^3_{-})^{2}}{2s}
	+i\frac{e\mathcal{B}}{4}\cot\!\left(\frac{e\mathcal{B}s}{2}\right)\bm x_{-\perp}^{2}
	+i\bm q^2
	\right]\,.\nonumber	
\end{align}
Here, we drop the phase term proportional to $V^2$ because it gives a small contribution. The obtained expression describes the propagation of a scalar particle in the combined fields, where the atomic potential is evaluated along the classical trajectory in the magnetic field.

For a pure magnetic field the result can be extended to an arbitrary angle $\theta$
between the classical velocity and the field.  Performing the Gaussian integrals we obtain
\begin{align}\label{eq:G_B_arb}
	&G(x_2,x_1)=i\,e^{i\Phi(x_{2\perp_B},x_{1\perp_B})}
	\int_{0}^{\infty}\frac{ds}{(2\pi)^3s^2}\int d \bm q \,
	\frac{e\mathcal{B}s/2}{\sin(e\mathcal{B}s/2)}\,
	\\
	&\times
	\exp\left[-i\frac{sm^{2}}{2}-i\frac{((x^0_{-})^{2}-(x_{-\parallel_B})^{2})}{2s}
	+i\frac{e\mathcal{B}}{4}\cot\!\left(\frac{e\mathcal{B}s}{2}\right)\bm x_{-\perp_B}^{2}
	+i\bm q^2
	\right]\,,\nonumber	\\
	&\times \exp\left[
	-i\frac{x_2^0-x_1^0}{s}\int_{0}^{s}d\tau V\left(\bm y_{cl}(\tau)+c(s)\bm q+\big[c_1(s)-c(s)\big](\tilde{\bm e}\cdot\bm q)\tilde{\bm e}\right)\right]
\end{align}
where
\begin{align}\label{eq:B_arb_c1}
	c_1^2(s)=c^2(s)\cos^2\theta+\frac{2\tau_0(s-\tau_0)}{s}\,\sin^2\theta\,,\qquad
	\tilde{\bm e}=\frac{\bm n_B-\bm\nu(\tau_0)
		\big(\bm\nu(\tau_0)\cdot\bm n_B\big)}{\sin\theta}\,.
\end{align}
Here $\bm n_B=\bm B/|\bm B|$, $\tilde{\bm e}$ is the unit vector lying in the plane spanned by $\bm\nu(\tau_0)$ and $\bm n_B$ and perpendicular to $\bm\nu(\tau_0)$, the vector $\bm q$ is understood as a vector in the plane transverse to $\bm\nu(\tau_0)$, and $\theta$ is the angle between $\bm\nu(\tau_0)$ and $\bm n_B$. The subscripts $\perp_B$ and $\parallel_B$  mean transverse and parallel with respect to $\bm B$. For a pure magnetic field the angle $\theta$ is constant along the classical trajectory, since the velocity components along the field and transverse to it are separately conserved. 

The classical trajectory is given by Eq.~\eqref{eq:ycl_perp}, with the perpendicular components understood with respect to $\bm B$, and by $y_{cl,\parallel_B}(\tau)=x_{1\parallel_B}+\frac{\tau}{s}(x_{2\parallel_B}-x_{1\parallel_B})$.

This extension is subject to the same no-return condition~\eqref{eq:no_return}. For the helical trajectory it requires the longitudinal drift over one cyclotron period, $\Delta\tau=2\pi/(e\mathcal B)$, to be large compared to the Compton wavelength, that is, $\frac{2\pi|\bm y'_{cl}|\,|\cos\theta|}{e\mathcal B}\gg\frac{1}{m}$. Thus all angles are allowed except a narrow region of order $m/\varepsilon$ around $\theta=\pi/2$. The Green's function at fixed energy is obtained from Eq.~\eqref{eq:G_B_arb} by the same Fourier transform over $x_2^0-x_1^0$ as in the small-angle case.

The singularities of the integrand at $e\mathcal{B}s/2=\pi n$, $n=1,2,\ldots$, are caustics of the transverse motion. They do not correspond to additional encounters with the atom, because the longitudinal coordinate continues to grow. The single-point reduction of the fluctuations is a local statement in the proper time and remains valid in the vicinity of these points. In Eqs.~\eqref{eq:G_energy_B_s} and~\eqref{eq:G_B_arb} we therefore keep the proper-time integral unexpanded, with the prescription $s\to s-i0$, which treats the transverse caustics exactly and encodes the Landau-level structure.

\section{Green's function for a spin-1/2 particle}\label{sec:spinor}

In this section we show that the results obtained above for a scalar particle can be extended to a particle with spin $1/2$. The spin degrees of freedom do not modify the spin-independent part of the path integral: they enter as a matrix factor along the trajectory. This factor generates new spinor structures in the Green's function, which must be treated with care. In particular, although the atomic contribution to the spin factor is suppressed by $Z\alpha\,m/\varepsilon_{char}$, it can be equally important in applications: in the matrix elements of the basic QED processes the leading term is itself suppressed, so that both contributions are of the same order~\cite{KLM2014,KLM2015}.

We define the Dirac Green's function as the solution of
\begin{align}\label{eq:Dirac_def}
	(\hat\Pi_{x_2}-m)\,S(x_2,x_1)=\delta^{(4)}(x_2-x_1)\,,\qquad
	\Pi_\mu=p_\mu-eA_\mu\,,\quad p_\mu=i\partial_\mu\,.
\end{align}
Note that in this section $A_\mu$ and $F_{\mu\nu}$ denote the total electromagnetic field, that is, the sum of the background and atomic contributions, whereas in Secs.~\ref{sec:general}--\ref{sec:const_atomic} they referred to the background field alone. Accordingly, we write $F_{\mu\nu}=F^{(0)}_{\mu\nu}+F^{(V)}_{\mu\nu}$, where $F^{(0)}_{\mu\nu}$ is the background field strength and $eF^{(V)}_{0i}=-\partial_i V$ corresponds to the atomic potential. Using $[\Pi_\mu,\Pi_\nu]=-ieF_{\mu\nu}$ and
$\Sigma^{\mu\nu}=\frac{i}{2}[\gamma^\mu,\gamma^\nu]$, one obtains
\begin{align}\label{eq:squared_Dirac}
	\hat\Pi^{\,2}=\Pi^2-\frac{e}{2}\,\Sigma^{\mu\nu}F_{\mu\nu}\,.
\end{align}
Therefore the Dirac Green's function can be expressed through the Green's function of the squared Dirac operator,
\begin{align}\label{eq:S_via_G2}
	S(x_2,x_1)=(\hat\Pi_{x_2}+m)\,\mathcal{G}(x_2,x_1)\,,
\end{align}
where $\mathcal{G}$ is a matrix in spinor space satisfying
\begin{align}\label{eq:G2_eq}
	\left[\Pi^2-m^2-\frac{e}{2}\,\Sigma^{\mu\nu}F_{\mu\nu}\right]\mathcal{G}(x_2,x_1)
	=\delta^{(4)}(x_2-x_1)\,.
\end{align}
Equation~\eqref{eq:G2_eq} differs from the scalar equation only by the Pauli term $\frac{e}{2}\Sigma^{\mu\nu}F_{\mu\nu}$. For $F_{\mu\nu}\to0$ one has $\mathcal{G}\to G\cdot\mathbb{1}$, where $G$ is the scalar Green's function.

Since the additional term does not contain derivatives, the proper-time representation of $\mathcal{G}$ has the same spin-independent action $S[y]$~\eqref{eq:G_combined} and the same measure as in the scalar case,
\begin{align}\label{eq:G2_path_integral}
	\mathcal{G}(x_2,x_1)=\int_{0}^{\infty}\frac{ds}{2i}\int\mathcal{D}^{4}y\,
	\exp\left\{iS[y]\right\}\,\Phi[y]\,,
\end{align}
where the spin factor is the path-ordered exponential
\begin{align}\label{eq:spin_factor}
	\Phi[y]=\mathcal{P}\exp\left[-\frac{ie}{4}\int_{0}^{s}d\tau\,
	\Sigma^{\mu\nu}F_{\mu\nu}(y(\tau))\right]\,.
\end{align}
The path ordering is necessary because the matrices $\Sigma^{\mu\nu}F_{\mu\nu}$ taken at different points of the trajectory do not commute. Equivalently, the spin factor can be obtained by introducing Grassmann variables on the world line and integrating them out~\cite{schubert2001perturbative}; here we keep the matrix structure explicitly.

Let us now estimate the spin factor for the superposition of the atomic and background fields. For the atomic field $eF^{(V)}_{0i}=-\partial_i V$, the contribution is accumulated in the formation region, where the trajectory is nearly
straight, so that we can use the longitudinal coordinate $z$ along $\bm\nu(\tau_0)$ and
the impact parameter $b$:
\begin{align}\label{eq:spin_atomic_estimate}
	\left|\frac{e}{4}\int_{0}^{s}d\tau\,\Sigma^{\mu\nu}F^{(V)}_{\mu\nu}\right|
	\sim\int_{0}^{s}d\tau\,\left|\nabla V(\bm y_{cl}(\tau))\right|
	\sim\frac{s}{|\bm x_2-\bm x_1|}\int dz\,\frac{Z\alpha}{z^{2}+b^{2}}
	\sim\frac{Z\alpha}{\varepsilon_{char}b}\sim Z\alpha\,\frac{m}{\varepsilon_{char}}\ll1\,.
\end{align}

The atomic part of the spin factor is accumulated only in the small region $|\tau-\tau_0|\sim1/(m\varepsilon_{char})$. Over this region the background spin factor is close to unity, since $|eF^{(0)}|\,|\tau-\tau_0|\lesssim m/\varepsilon_{char}$ by Eq.~\eqref{eq:field_strength}. The atomic insertion can therefore be placed between the two background kernels. Over the whole trajectory, in contrast, $|eF^{(0)}|\,s/4\lesssim1$, so the background spin factor  is of order unity. Thus the spin factor has the following form:
\begin{align}\label{eq:spin_factor1}
	&\Phi[y]=\mathcal{P}\exp\left[-\frac{ie}{4}\int_{\tau_0}^{s}d\tau\,
	\Sigma^{\mu\nu}F^{(0)}_{\mu\nu}(y(\tau))\right] 
	\exp\left[-\frac{ie}{4}\int_{0}^{s}d\tau\,
	\Sigma^{\mu\nu}F^{(V)}_{\mu\nu}(y(\tau))\right]\\
	&
	\times\mathcal{P}\exp\left[-\frac{ie}{4}\int_{0}^{\tau_0}d\tau\,
	\Sigma^{\mu\nu}F^{(0)}_{\mu\nu}(y(\tau))\right]\,.\nonumber
\end{align}
Repeating the derivation of Sec.~\ref{sec:general}, we obtain the spinor analogue of the general formula:
\begin{align}\label{eq:G_general_master_spin}
	&\mathcal{G}(x_2,x_1)=
	\int_{0}^{\infty}\frac{ds}{2i}\int d^{4}y_*\,
	\mathfrak{g}_0\!\left(x_2,y_*;s-\tau_0\right)\\
	&	\exp\left[-i\int_{0}^{s}d\tau\,\left({y'}_{cl}^0(\tau)-i\frac{\bm\alpha \cdot\bm \partial }{2\partial (\delta\bm y_{*\perp}) }\right)\,
	V(\bm y_{cl}(\tau)+\delta\bm y_{*\perp})\right]
	\mathfrak{g}_0\!\left(y_*,x_1;\tau_0\right)	\,,\nonumber
\end{align}
where $\mathfrak{g}_0$ is the proper-time kernel of the squared Dirac operator in the background field, which includes the corresponding spin factor. The atomic phase is inserted between the two kernels, which automatically provides the correct ordering of the matrix factors: the spin-dependent part of the atomic phase is localized at $\tau_0$, so that the background spin factors of the two intervals stand to the left and to the right of it. Note that the substitution is understood to first order in the spin term. The derivative with respect to $\delta\bm y_{*\perp}$ acts on the potential and gives $\bm\alpha\cdot\nabla_\perp V$. The longitudinal part of $\bm\alpha\cdot\nabla V$ is suppressed by an additional factor $m/\varepsilon_{char}$, by the same argument as in Sec.~\ref{sec:general}, and is therefore omitted.

It is instructive to consider the two backgrounds used above. For a constant field the
spin factor is an ordinary matrix exponential,
\begin{align}
	\Phi_0(\tau_2,\tau_1)=\mathcal{P}\exp\left[-\frac{ie}{4}\int_{\tau_1}^{\tau_2}d\tau\,
	\Sigma^{\mu\nu}F^{(0)}_{\mu\nu}(y(\tau))\right] =\exp\left[-\frac{ie (\tau_2-\tau_1)}{4}\,\Sigma^{\mu\nu}F^{(0)}_{\mu\nu}\right]\,,
\end{align}
which, together with the scalar kernel of Appendix~\ref{app:const}, reproduces the known spinor Green's function in a constant field~\cite{Schwinger1951,ItzyksonZuber}. 

For a plane wave, $F^{(0)}_{\mu\nu}=k_\mu a_\nu-k_\nu a_\mu$ with $k^2=0$, $k\cdot a=0$, so   that $\Sigma^{\mu\nu}F^{(0)}_{\mu\nu}=2i\hat k\hat a$ is nilpotent and all products of the spin factors at different phases vanish. The path ordering becomes irrelevant and the exponential terminates at the first order,
\begin{align}
	\Phi_0(\tau_2,\tau_1)=1+\frac{e}{2\phi'_{cl}}\,\hat k
	\left[\hat A^{(0)}(\phi_{\tau_2})-\hat A^{(0)}(\phi_{\tau_1})\right]\,,
\end{align}
where $k^\mu=(1,\bm n)$ is the null vector of the wave, $\phi=k\cdot x$, $\phi_{\tau_{1,2}}=\phi_{cl}(\tau_{1,2})$, and $a_\mu=dA^{(0)}_\mu/d\phi$. It  is the standard spinor structure of the Volkov solutions. 

To summarize, at leading order in $m/\varepsilon_{char}$ the spinor quasiclassical Green's function is obtained from the scalar one by inserting the background spin factors into the proper-time kernels, by including the atomic spin term in the phase as in Eq.~\eqref{eq:G_general_master_spin}, and by applying the operator $(\hat\Pi_{x_2}+m)$ at the endpoint. The single-point fluctuation reduction applies in the same way as in the scalar case, so all the results of Secs.~\ref{sec:QQGF_atom}--\ref{sec:const_atomic} carry over to a spin-1/2 particle.

\section{Conclusion}\label{sec:concl}

In this paper, we have developed the trajectory-localized fluctuation method within the path integral framework and applied it to the quasiclassical Green's function  of a high-energy charged particle in an atomic potential, both alone and in combination with a smooth electromagnetic background. We have shown that the essential quantum fluctuations in the argument of the atomic potential are confined to a small region near the point of closest approach, so that they can be replaced by the fluctuation at a single  point. This allows one to reduce the functional integral to a finite-dimensional one, with the potential evaluated along the classical trajectory shifted by a single transverse fluctuation.

Since the method applies primarily to the coordinate dependence of the Green's function, we first consider the scalar particle. For the pure atomic potential, the method reproduces the known quasiclassical Green's function~\eqref{eq:G_atom_res}. For the superposition with a smooth background, we obtain the general formula~\eqref{eq:G_general_master}, in which the atomic potential is evaluated along the classical trajectory in the background field. The formula is valid for high-energy kinematics under the conditions~\eqref{eq:kin_arb}, \eqref{eq:field_strength} and~\eqref{eq:no_return}.

For the plane-wave laser field, we obtain the Green's function~\eqref{eq:G_laser_x_arb_new} for an arbitrary angle between the particle and the laser propagation direction. In the head-on kinematics it reduces to the known result of Refs.~\cite{DM12,DM14} at fixed light-cone energy. 

For the constant electromagnetic field, we obtain the covariant expression~\eqref{eq:G_const_x_fluct_cov} and the explicit result~\eqref{eq:G_B_E_parallel} for parallel electric and magnetic fields. For the pure magnetic field, the Green's function is obtained both for a small angle and for an arbitrary angle between the particle and the field, Eqs.~\eqref{eq:G_energy_B_s} and~\eqref{eq:G_B_arb}.

We have also shown that the same framework extends to a spin-$1/2$ particle. The spin degrees of freedom enter as an additional matrix factor along the trajectory, while the reduction of the fluctuations remains the same. The atomic part of the spin factor is of order $Z\alpha\,m/\varepsilon_{char}$ and is taken into account to first order, Eq.~\eqref{eq:G_general_master_spin}. It can be important in applications because in the matrix elements of the basic QED processes the leading term is itself suppressed. The background spin factor is of order unity and is kept exactly in the proper-time kernels; for the constant field and for the plane wave this reproduces the known spinor Green's functions~\cite{Schwinger1951,ItzyksonZuber}.

All results are exact in the atomic potential, to all orders in $Z\alpha$, and in the background field parameters, and hold at leading order in $m/\varepsilon_{char}$.  We assume that the background is smooth on the Compton scale and is less than or of the order of the critical Schwinger value; for the laser field we additionally assume $\xi\lesssim1$, so that the laser-induced transverse momentum remains of order $m$.  The general formula also assumes that the proper-time kernel of the background is known in closed form. 

The obtained Green's functions can be used for the calculation of the cross sections of various QED processes in the combined fields at high energies. The method developed here is of independent interest and can be applied to more complicated systems.

\section*{Acknowledgments}
We are grateful to V.O. Guba for the fruitful discussions.
\appendix
\section{Green's function in the laser field}\label{app:laser}
In this appendix, we derive the exact Volkov Green's function in the path integral formalism. This result is used in Section~\ref{sec:laser_atomic} for the derivation of the Green's function  in the superposition of laser and atomic fields. Let the laser field propagate along the  unit vector $\bm n$. The plane-wave vector potential is $\bm  A(t-\bm n\cdot\bm r)$, with the condition $\bm n\cdot\bm A=0$. It is convenient to use  the light-cone variables:
\begin{align}\label{eq:lightcone}
	\phi=y^0-\bm n\cdot\bm y\,,\qquad T=\frac{y^0+\bm n\cdot\bm y}{2}\,,\quad  \bm y_{n\perp}=\bm y-\bm n (\bm y\cdot\bm n),
\end{align}
for any four-vector $y$. In these variables, the four-vector square is $y^{2}=2\phi T-\bm y_{n\perp}^{2}$. The Green's function takes the form:
\begin{align}\label{eq:G_laser}
	G_0(x_2,x_1)\simeq\int_{0}^{\infty}ds\,\int\mathcal{D}^{4}y\,\exp\left[-i\frac{sm^{2}}{2}-\frac{i}{2}\int_{0}^{s}d\tau\left[2\phi'(\tau)T'(\tau)-{y'}^{2}_{n\perp}(\tau)\right]+ie\int_{0}^{s}d\tau\bm{y}'_{n\perp}(\tau)\cdot\bm{A}(\phi(\tau))\right]\,.
\end{align}
Here we assume the boundary conditions:
\begin{align}\label{eq:laser_boundary}
	y^\mu(0)=x^\mu_1\,,\quad y^\mu(s)=x^\mu_2\,,
\end{align}
thus $T(0)=T_1$, $T(s)=T_2$, $\phi(0)=\phi_1$, $\phi(s)=\phi_2$, $\bm y_{n\perp}(0)=\bm x_{1n\perp}$, $\bm y_{n\perp}(s)=\bm x_{2n\perp}$.

The integrals over $\bm y_{n\perp}(\tau)$ are Gaussian and can be evaluated exactly. Completing the square:
\begin{align}
	&\int\mathcal{D}^{2}y_{n\perp}\,
	\exp\left[\frac{i}{2}\int_{0}^{s}d\tau\,{y'}_{n\perp}^{2}(\tau)
	+ie\int_{0}^{s}d\tau\,\bm y_{n\perp}'(\tau)\cdot\bm  A(\phi(\tau))\right]\nonumber\\
	&\simeq\exp\left[\frac{i}{2s}\left[\bm y_{n\perp}(s)-\bm y_{n\perp}(0)+\int_{0}^{s}d\tau\,e\bm  A(\phi(\tau))\right]^{2}
	-\frac{i}{2}\int_{0}^{s}d\tau\,e^2\bm  A^{2}(\phi(\tau))\right]\,.
\end{align}

The integral over $T(\tau)$ produces a $\delta$-function enforcing $\phi''(\tau)=0$
\begin{align}
	\int\mathcal{D} T \exp\left[-i\int_{0}^{s}d\tau\phi'(\tau)T'(\tau)\right]\simeq\delta\left(\phi''(\tau)\right)\exp\left[-i\left[\phi'(s)T(s)-\phi'(0)T(0)\right]\right]\,,
\end{align}
which means that $\phi'(\tau)=(\phi(s)-\phi(0))/s$ is a constant. Using the boundary conditions~\eqref{eq:laser_boundary}, we obtain:
\begin{align}
	\int\mathcal{D}\phi\,\delta\left(\phi''(\tau)\right)\,
	\exp\left[-i\left[\phi'(s)T(s)-\phi'(0)T(0)\right]\right]
	=\exp\left[-i\frac{(\phi_2-\phi_1)(T_2-T_1)}{s}\right]\,.
\end{align}

Combining the previous results with the fact that $\phi(\tau)=\phi_1+\frac{\phi_2-\phi_1}{s}\tau$, we obtain:
\begin{align}\label{eq:G_laser_x_app}
	&G_0(x_2,x_1)=-\frac{1}{2}\int_{0}^{\infty}\frac{ds}{(2\pi s)^2}\,\\
	\times&\exp\left[-i\frac{sm^{2}}{2}-i\frac{(\phi_2-\phi_1)(T_2-T_1)}{s}
	+\frac{i}{2s}\left[\bm x_{2n\perp}-\bm x_{1n\perp}+s\int_{0}^{1}du\,e\bm  A(\phi_{u})\right]^{2}
	-i\frac{s}{2}\int_{0}^{1}du\,e^2\bm A^{2}(\phi_{u})
	\right]\,,\nonumber
\end{align}
where we perform the change of variables $u=\tau/s$ and use the notation $\phi_{u}=\phi_1+u(\phi_2-\phi_1)$.

Since all path-integral manipulations above produce no prefactor depending on the vector potential, the prefactor coincides with that of the free-particle Green's function.
The result coincides with the standard Volkov Green's function for a scalar particle in a plane-wave background~\cite{ItzyksonZuber}. 

The Green's function at fixed light-cone energy  $\varepsilon_{T}$ is obtained by the Fourier transform over $\Delta T=T_2-T_1$. The integral over $\Delta T$ produces the $\delta$-function $\delta(\varepsilon_{T}-(\phi_{2}-\phi_{1})/s)$, which fixes $s=(\phi_{2}-\phi_{1})/\varepsilon_{T}$. The result is:

\begin{align}\label{eq:G_laser_final_app}
	&G_0(\bm x_{2n\perp},\phi_2;\bm x_{1n\perp},\phi_1|\varepsilon_{T})
	=\frac{-1}{4\pi(\phi_2-\phi_1)}\\
	&\times\exp\left[-i\frac{\phi_2-\phi_1}{2\varepsilon_T}\left(m^2
	-\left[\varepsilon_T \frac{\bm x_{2n\perp}-\bm x_{1n\perp}}{\phi_2-\phi_1}+\int_{0}^{1}du\,e\bm  A(\phi_{u})\right]^{2}
	+\int_{0}^{1}du\,e^2\bm A^{2}(\phi_{u})\,\right)\right]\,.\nonumber
\end{align}
The combination $\varepsilon_{T} \frac{\bm x_{2n\perp}-\bm x_{1n\perp}}{\phi_{2}-\phi_{1}}+\int_{0}^{1}du\,e\bm  A(\phi_{u})$  can be interpreted as the effective transverse momentum of the particle in the laser field.

We emphasize that the derivation above is exact and does not assume any specific kinematics of the charged particle. The boundary conditions $x_1$ and $x_2$ in Eq.~\eqref{eq:laser_boundary} are arbitrary, and the results~\eqref{eq:G_laser_x_app} and~\eqref{eq:G_laser_final_app} are valid for any angle between the particle momentum and the laser propagation direction. The light-cone variables $\phi$ and $T$ are defined relative to the laser propagation axis, but the particle endpoints $x_1$ and $x_2$ need not satisfy any special kinematic constraints.

For completeness, we write here the classical trajectory of the charged particle in the presence of the laser field:
\begin{align}\label{eq:y_cl}
	\bm y_{cl,n\perp}(\tau)=&\bm x_{1n\perp}+\frac{\tau}{s} 
	\left[\bm x_{2n\perp}-\bm x_{1n\perp}+\int_0^s d\tau'\,e\bm A(\phi_{cl}(\tau'))\right] 
	-\int_0^\tau d\tau'\,e\bm A(\phi_{cl}(\tau'))\,,\\
	\phi_{cl}(\tau)=&\phi_1+\frac{\phi_2-\phi_1}{s}\,\tau\,,\nonumber \\
		T_{cl}(\tau)=&T_1+\frac{\tau}{s}(T_2-T_1) +\frac{1}{2\phi_{cl}'}\left[\int_0^\tau d\tau'\,\bm y_{cl,n\perp}'^{\,2}(\tau')
	-\frac{\tau}{s}\int_0^s d\tau'\,\bm y_{cl,n\perp}'^{\,2}(\tau')	\right]\,.\nonumber
\end{align}

\section{Green's function in a constant electromagnetic field}\label{app:const}

In this appendix we consider the exact Green's function of a scalar particle in a constant homogeneous electromagnetic field, which was first obtained by Schwinger in~\cite{Schwinger1951} and can be found in the textbook~\cite{ItzyksonZuber}. For such a case the phase is a quadratic function, and the path integral is Gaussian. Although the result is standard, we present a brief derivation for completeness.

We use the symmetric (Fock--Schwinger) gauge $	A_\mu(y)=-\frac{1}{2}F_{\mu\nu}y^\nu$, which reproduces $F_{\mu\nu}=\partial_\mu A_\nu-\partial_\nu A_\mu$ and satisfies $\partial^\mu A_\mu=0$. The action~\eqref{eq:S0} is quadratic:
\begin{align}
	S_0[y]=-\frac{sm^2}{2}-\frac{1}{2}\int_{0}^{s}d\tau\,y'^2(\tau)
	+\frac{e}{2}\int_{0}^{s}d\tau\,y'^\mu(\tau)F_{\mu\nu}y^\nu(\tau)\,,
\end{align}
and the stationary trajectory obeys the Lorentz equation~\eqref{eq:Lorentz}, $y_{cl}''^\mu=eF^{\mu}_{\ \nu}y_{cl}'^\nu$ with 
boundary conditions $y_{cl}(0)=x_1$, $y_{cl}(s)=x_2$. Since $F$ is constant, this equation can be trivially  solved:
\begin{align}\label{eq:ycl_const}
	y_{cl}(\tau)&=x_1+\frac{e^{eF\tau}-1}{e^{eFs}-1}\,x_-\,,
\end{align}
where $x_\pm=x_2\pm x_1$.

The classical action $S_0[y_{cl}]$ has the following form:
\begin{align}
	S_0[y_{cl}]=-\frac{sm^2}{2}-\frac{1}{2}\Big[y_{cl}\cdot y'_{cl}\Big]_{0}^{s}=-\frac{sm^2}{2}
	-\frac{1}{4}\,x_-^\mu\Big[eF\coth\!\big(eFs/2\big)\Big]_{\mu\nu}x_-^\nu
	+\Phi(x_2,x_1)\,,
\end{align}
where the antisymmetric part of the matrix drops out of the quadratic form, and
\begin{align}\label{eq:Schwinger_phase}
	\Phi(x_2,x_1)=-e\int_{x_1}^{x_2}A_\mu\,dy^\mu
	=\frac{e}{4}\,x_-^\mu F_{\mu\nu}x_+^\nu\,,
\end{align}
is the Schwinger phase, taken along the straight line connecting the endpoints. It carries the entire gauge dependence of the Green's function.

We pass from integration over $y(\tau)$ to integration over the fluctuations $\delta y(\tau)=y(\tau)-y_{cl}(\tau)$, with boundary conditions $\delta y(0)=\delta y(s)=0$. Since the classical trajectory corresponds to the extremum of the action $S_0$ and $S_0$ is a quadratic function, the action $S_0[y_{cl}+\delta y]$ can be written as a sum $S_0[y_{cl}+\delta y]=S_0[y_{cl}]+\delta S_0[\delta y]$, where $\delta S_0[y]=S_0[y]+\frac{sm^2}{2}$. Thus, the path integral over quantum fluctuations $\delta y(\tau)$ corresponds to the Green's function prefactor. This prefactor can be written as a functional determinant, which can be computed using the Gelfand--Yaglom method~\cite{GelfandYaglom60}. We write the equation $\ddot Y=eF\dot Y$ with $Y(0)=0$, $\dot Y(0)=\mathbb{1}$, whose solution is $Y(\tau)=(e^{eF\tau}-1)(eF)^{-1}$. This gives the field-dependent factor
\begin{align}\label{eq:det_const}
	\mathcal{D}(s)=\det{}^{-1/2}\!\left[\frac{\sinh(eFs/2)}{eFs/2}\right],
\end{align}
the determinant being taken over the four Lorentz indices. Since the determinant of a tensor of type $(1,1)$ is a Lorentz invariant, $\mathcal{D}(s)$ can be expressed through the invariants $\mathcal{E}$ and $\mathcal{B}$ defined below in Eq.~\eqref{eq:eff_fields} as
\begin{align}\label{eq:det_const_explicit}
	\mathcal{D}(s)=\frac{e\mathcal{E}s/2}{\sinh(e\mathcal{E}s/2)}\,
	\frac{e\mathcal{B}s/2}{\sin(e\mathcal{B}s/2)}\,,
\end{align}
which is valid in any reference frame. The remaining, field-independent normalization is the free one, in accordance with Eq.~\eqref{eq:Green_start}. The exact Green's function is therefore
\begin{align}\label{eq:G_const_app}
	G_0(x_2,x_1)=\int_{0}^{\infty}\frac{ds}{2(2\pi i s)^{2}}\,
	\mathcal{D}(s)\,e^{i\Phi(x_2,x_1)}
	\exp\left\{-i\frac{sm^{2}}{2}
	-\frac{i}{4}x_-^\mu\Big[eF\coth\!\big(eFs/2\big)\Big]_{\mu\nu}x_-^\nu\right\}\,.
\end{align}

Both the determinant and the quadratic form in the exponent are expressed through the two invariants of the constant field:
\begin{align}
	\mathcal{F}=\frac{1}{4}F_{\mu\nu}F^{\mu\nu}=\frac{\bm B^2-\bm E^2}{2}\,,
	\qquad
	\mathcal{G}=\frac{1}{4}F_{\mu\nu}\tilde F^{\mu\nu}\,,
\end{align}
where $\tilde F^{\mu\nu}=\frac{1}{2}\epsilon^{\mu\nu\alpha\beta}F_{\alpha\beta}$ and $\epsilon^{\mu\nu\alpha\beta}$ is a four-dimensional Levi-Civita tensor. It is convenient to introduce the invariant field strengths
\begin{align}\label{eq:eff_fields}
	\mathcal{B}=\sqrt{\sqrt{\mathcal{F}^2+\mathcal{G}^2}+\mathcal{F}}\,,
	\qquad
	\mathcal{E}=\sqrt{\sqrt{\mathcal{F}^2+\mathcal{G}^2}-\mathcal{F}}\,,
\end{align}
so that $\mathcal{B}^2-\mathcal{E}^2=\bm B^2-\bm E^2$ and $\mathcal{B}\mathcal{E}=|\bm E\cdot\bm B|$. 

In the frame where $\bm E\parallel\bm B$ one has $\mathcal{B}=|\bm B|$ and $\mathcal{E}=|\bm E|$; both are nonzero as long as $\mathcal{G}\neq0$. The eigenvalues of $F^{\mu}_{\ \nu}$ are $\pm\mathcal{E}$ and $\pm i\mathcal{B}$. Let $\bm n_B=\bm B/|\bm B|$. For an arbitrary vector $\bm x$ we denote $\bm x_{\perp_B}=\bm x-\bm n_B(\bm n_B\cdot\bm x)$ and $x_{\parallel_B}=\bm x\cdot \bm n_B$.

In this frame, Eq.~\eqref{eq:G_const_app} becomes:
\begin{align}\label{eq:G_const_app_EB}
	&G_0(x_2,x_1)=e^{i\Phi(x_2,x_1)}\int_{0}^{\infty}\frac{ds}{2(2\pi i s)^{2}}\,
	\frac{e\mathcal{E}s/2}{\sinh(e\mathcal{E}s/2)}\,
	\frac{e\mathcal{B}s/2}{\sin(e\mathcal{B}s/2)}\,\exp\left\{i\mathcal{S}_0\right\},\\
	&\mathcal{S}_0=-\frac{sm^{2}}{2}
	-\frac{e\mathcal{E}}{4}\coth\!\left(\frac{e\mathcal{E}s}{2}\right)((x^0_{-})^{2}-(x_{-\parallel_B})^{2})
	+\frac{e\mathcal{B}}{4}\cot\!\left(\frac{e\mathcal{B}s}{2}\right)\bm x_{-\perp_B}^{2}\,.\nonumber
\end{align}
The magnetic field produces the oscillating functions $\sin$ and $\cot$, reflecting the Landau-level structure, while the electric field produces the hyperbolic functions $\sinh$ and $\coth$, related to pair creation. For $\mathcal{E}\to0$ ($\mathcal{B}\to0$) the corresponding factor tends to unity and $\coth$ ($\cot$) term reduces to its free form, reproducing the pure magnetic (electric) case; for $F\to0$ Eq.~\eqref{eq:G_const_app_EB} reduces to the free propagator.

Finally, we note that the integrand of Eq.~\eqref{eq:G_const_app_EB} has caustics at $e\mathcal{B}s/2=\pi n$, where $n=1,2,\ldots$, at which multiple classical trajectories connecting the same endpoints coexist, differing by  an integer number of complete orbits. The exact treatment of caustics corresponds to the prescription $s\to s-i0$~\cite{kleinert2006path}, which provides the correct pole circumvention and automatically sums over all these trajectories with proper Maslov phases~\cite{MaslovFedoriuk1981}.

\bibliography{path}
\end{document}